\documentclass[a4paper]{article}
\pdfoutput=1
\usepackage{graphicx}
\usepackage{amsmath,amssymb}
\usepackage{array,blindtext}

\newcommand\beq{\begin{equation}}
\newcommand\eeq{\end{equation}}
\newcommand\bea{\begin{eqnarray}}
\newcommand\eea{\end{eqnarray}}

\begin{document}
\title{SU(2) Lattice Gauge Theory- Local Dynamics on Non-intersecting Electric flux Loops}
\author{Ramesh Anishetty\footnote{ramesha@imsc.res.in} ~~and ~ Indrakshi Raychowdhury\footnote{indrakshi@imsc.res.in}\\
The Institute of Mathematical Sciences, \\\vspace{0.7cm} CIT-Campus, Taramani,
Chennai, India}
\maketitle
\begin{abstract}
We use Schwinger Bosons as prepotentials for lattice gauge theory to define  local linking operators and calculate their action on linking states for $2+1$ dimensional SU(2) lattice gauge theory. We develop a diagrammatic technique and associate a set of (lattice Feynman) rules  to compute the entire loop dynamics diagrammatically.  The physical loop space is shown to contain only non-intersecting loop configurations after solving the Mandelstam constraint. The smallest plaquette loops are contained in the physical loop space and other configurations are generated by the action of a set of fusion operators  on this basic loop states enabling one to charaterize any arbitrary loop by the basic plaquette together with the fusion variables. Consequently, the full Kogut-Susskind Hamiltonian and the dynamics of all possible non-intersecting physical loops are formulated in terms of these fusion variables.
\end{abstract}
\section{Introduction}

Lattice gauge theories, originally defined \cite{wil} within the Euclidean framework has  found profound applicability for performing numerical computations using Monte Carlo simulation. The Hamiltonian approach \cite{kogut}, although much less studied, has several important advantages over the Euclidean one. 
Both the Hamiltonian and path integral approach of lattice gauge theories are mostly studied in the strong coupling limit, albeit the physical/continuum limit exists at weak coupling.  Moreover, the most economic and physical description of any gauge theory can only be  in terms of gauge invariant degrees of freedom. Reformulation of gauge theories in terms of gauge invariant Wilson loops and strings carrying fluxes is an old problem in physics \cite{loops,mm}. Formulation of gauge field theories on lattice \cite{wil} is indeed an important step towards the loop formulation as here one directly works with the link variables or holonomies (instead of the gauge field for continuum theories) which are gauge-covariant objects and are the fundamental building blocks of gauge invariant Wilson loops. However, the gauge invariant wilson loops and strings form a over-complete basis for the physical Hilbert space of the theory.
Mandelstam constraints \cite{mans} indeed restricts the overcomplete Wilson loops to minimal loops which are also sufficiently complete to describe the physical Hilbert space. 
 But that is not a trivial task mostly because of the nonlocality of the Wilson loop states and their dynamics.  This problem becomes more and more tedious when one approaches the weak coupling limit of lattice gauge theory, where all possible loops of arbitrary shapes and sizes start contributing. However, in the context of duality transformation \cite{ramesh}, the electric flux loop  and their dynamics has been shown to be manifestly local in the continuum limit even for non-Abelian lattice gauge theories. Moreover,
 a recent development in the formulation of Hamiltonian lattice gauge theory, namely the prepotential formulation \cite{mm,pp} has shown a way to get rid of the problem of nonlocality and proliferation of loop states for any SU(N) gauge theory in arbitrary dimensions. 
 
The prepotential formulation is basically a reformulation of Hamiltonian lattice gauge theory in terms of SU(N) Schwinger Bosons in which the loop operators and loop states are defined locally at each site which cuts down the level of complications to a great extent. The Mandelstam constraints are also local in this formulation which one can solve  to find exact and local loop basis at each site. Thus this new local description of lattice gauge theory seems to provide the best framework for any practical computation in the field of lattice gauge theory. Besides strong coupling calculations the weak coupling regime becomes much more amenable and easy to handle in terms of prepotentials.

Using the Schwinger Boson representation of the gauge group  at each lattice site, the original Kogut Susskind Hamiltonian \cite{kogut} and its canonical conjugate variables are reconstructed. In terms of Schwinger Bosons, the non-Abelian gauge group becomes ultra local at each site and the fluxes along neighbouring sites flow following the new Abelian constraint, which  is easy to handle. However, the full Hamiltonian, even in terms of local gauge invariant operator is complicated enough while acting on an arbitrary loop state. In this work, exploiting the local description of loops in terms of Schwinger Bosons, we calculate all possible action of local gauge invariant operators on any local gauge invariant state of the theory with explicit realization for SU(2) lattice gauge theory defined on $2+1 $ dimensional lattice. Moreover, to realize the complicated actions and to perform computations (both analytical and numerical) easily we develop a diagrammatic calculational technique. We describe the local gauge invariant state as well as the actions of the gauge invariant operators on those states by diagrams. Each diagram denotes the states together with a numerical coefficient, which can be read off from it by a set  `lattice Feynman rules'. We utilize this diagrammatic technique to compute the action of full Kogut-Susskind Hamiltonian within loop states which is again expressed diagrammatically. Moreover, we improve the loop descriptions given in terms of local linking numbers in prepotential formulation to a description in terms of fusion variables. The Abelian Gauss laws are solved by these fusion variables by construction.  The electric part of the Hamiltonian is simple in terms of the fusion variables, which counts the units of flux flowing throughout the lattice and becomes dominant in the strong coupling limit. The magnetic part of the Hamiltonian which is dominant in the  weak coupling regime of the theory is quite complicated but have been written down entirely in terms of the shift operators corresponding to fusion variables. Both the diagrammatic representation as well as analytic expression is given.

The plan of the paper is as follows: we start with a brief review of the prepotential formulation and relate it to the Kogut-Susskind Hamiltonian formulation in section 2. In section 3, we discuss all possible loop operators in prepotential formulation defined locally at each site, and calculate their action individually on any loop state characterized by prepotential linking numbers. In this section we develop the diagrammatic technique to handle loops. Next in section 4, we shift from linking numbers to fusion variables to characterize any arbitrary loop states within the theory. We also introduce the shift operators corresponding to fusion variables which are responsible for loop dynamics. The associated constraints on the states characterized by fusion quantum numbers are also discussed which are there to define the loop states with only physical degrees of freedom. In section 5, we calculate the action of the full Kogut-Susskind Hamiltonian in terms of diagrams as well as the fusion variables. In section 6 we briefly illustrate how to compute strong coupling perturbation expansion within our formulation and compare our results for first few orders with available results. Finally we summarize our results in section 6 and also discuss the future directions.

\section{Prepotential Formulation: A Brief Review}
The prepotential formulation of lattice gauge theory \cite{pp} provides us with a platform to work with gauge invariant operators and states defined locally at each site of the lattice. We briefly review this particular formulation in this section for the sake of completeness. Note that, we keep ourselves confined to the  gauge group SU(2) and 2+1 dimensional lattice in this work, although each of these ideas can be generalized to arbitrary gauge group and arbitrary dimensions as well. 

 In  Kogut-Susskind \cite{kogut} formulation, the canonical conjugate variables in the theory are color electric fields $E^{\mathrm a}_{L/R}(x,e_i)$ defined at each site $x$, for $a=1,2,3$ and the $L/R$ denotes that the left electric field is located at the starting end of the link starting from $x$ along $e_i$ and $R$ denotes the electric field attached at the ending point terminating at $x+e_i$. The link operator $U(x,e_i)$'s are defined on a link originating from site  $x$ along $e_i$ direction. The Hamiltonian of the theory is given by,
\bea
\hskip -1.2cm H =g^2\sum_{x}\sum_{\mathrm a=1}^{3}E^{\mathrm a}(x,e_i) E^{\mathrm a}(x,e_i) 
- \frac{1}{g^2} \sum_{\mbox{plaquette}} Tr \left(U_{\mbox{plaquette}} + U^{\dagger}_{\mbox{plaquette}}\right)
\label{ham}
\eea
where, $g^2$ is the coupling constant.
In (\ref{ham}),
$
U_{\mbox{plaquette}}=U(x,e_1)U(x+e_1,e_2)U^{\dagger}(x+e_1+e_2,e_1)U^{\dagger}(x+e_2,e_1)$ is product over links around the smallest closed loop on a lattice, i.e a plaquette and 
$\mathrm a (=1,2,3)$ is the color index for SU(2). Note that, for SU(2) case, $ Tr U_{\mbox{plaquette}}= Tr U^{\dagger}_{\mbox{plaquette}}$.

The canonical conjugate variables, namely the color electric fields and the link operators  satisfy the commutation relation:
\bea
\left[E_L^{\mathrm a}(x,e_i),U^{\alpha}{}_{\beta}(x,e_i)\right] = - \left(\frac{\sigma^{\mathrm a}}{2}U(x,e_i)\right)^{\alpha}{}_{\beta},~~ 
~~~\left[E_R^{\mathrm a}(x+e_i),U^{\alpha}_{\beta}(x,e_i)\right] = \left(U(x,e_i) \frac{\sigma^{\mathrm a}}{2} \right)^{\alpha}{}_{\beta}. 
\label{ccr} 
\eea 
In (\ref{ccr}), $\frac{\sigma^{\mathrm a}}{2}$ are  the Pauli matrices, satisfying: $[\frac{\sigma^{\mathrm a}}{2},\frac{\sigma^{\mathrm b}}{2}]=i\epsilon^{\mathrm{abc}}\frac{\sigma^{\mathrm c}}{2}$. The left and right electric fields are generators of the gauge transformation and hence follow SU(2) algebra:
\bea
\label{eec}
[E_L^{\mathrm a}(x,e_i),E_L^{\mathrm b}(x,e_i)] &=& i\epsilon_{\mathrm{abc}}E_L^{\mathrm c}(x,e_i),\nonumber \\
\left[E_R^{\mathrm a}(x,e_i),E_R^{\mathrm b}(x,e_i)\right] &=& i\epsilon_{\mathrm{abc}}E_R^{\mathrm c}(x,e_i),\\
\left[E_L^{\mathrm a}(x,e_i),E_R^{\mathrm b}(x,e_i)\right]&=& 0.\nonumber
\eea
Note that the left and right generators $E^\mathrm a_L(x,e_i)$ and $E^\mathrm a_R(x+e_i,e_i)$ on the link $(x,e_i)$  are the 
parallel transport of each other, i.e 
$ E_R(x+e_i,e_i)= - U^{\dagger}(x,e_i) E_L(x,e_i) U(x,e_i)
$, implying,
\bea
\label{E2=e2}
\sum_{{\mathrm a}=1}^{3}E^{\mathrm a}(x,e_i)E^{\mathrm a}(x,e_i) \equiv 
\sum_{{\mathrm a}=1}^{3}E_L^{\mathrm a}(x,e_i)E_L^{\mathrm a}(x,e_i)
=\sum_{{\mathrm a}=1}^{3}E_R^{\mathrm a}(x+e_i,e_i)E_R^{\mathrm a}(x+e_i,e_i).
\eea
Hence the electric part of the Hamiltonian (\ref{ham}) contains either of the electric fields and we choose it to be the left electric field.
Under gauge transformation, the left electric field and 
the link operator transforms as:
\bea
\label{eugt}
U(x,e_i)\rightarrow \Lambda(x)U(x,e_i)\Lambda^{\dagger}(x+e_i), \hskip 4cm 
\nonumber \\ 
E_L(x,e_i)\rightarrow \Lambda(x,e_i)E_L(x,e_i)\Lambda^{\dagger}(x),~~
E_R(x+e_i,e_i)\rightarrow \Lambda(x+e_i)E_R(x+e_i,e_i)\Lambda^{\dagger}(x+e_i).
\eea
Also note that, from (\ref{eugt}), the $SU(2)$ Gauss law 
constraint at every lattice site $n$ is
\bea
\label{gl}
\hskip 2cm G(n) = \sum_{i=1}^{d}\Big( E_L^{\mathrm a}(x,e_i) + E_R^{\mathrm a}(x+e_i,e_i)\Big)=0, \forall x. 
\eea
In the next subsection we briefly review how the SU(2) Hamiltonian lattice gauge theory is reformulated in terms of prepotentials.

\subsection{Schwinger Bosons:}
Instead of associating electric fields and link operators to each link of the lattice as discussed before, let us associate a set of Harmonic oscillator doublets acting as Schwinger Bosons $a_{\alpha}(x,e_i;l)$ and $a_{\alpha}^{\dagger}(x,e_i;l)$ with $l= L,R, \alpha =1,2$. We call these oscillators as prepotentials since the electric field operators as well as the link operators can be reconstructed solely in terms of these.
\begin{figure}[h]
\begin{center}
\includegraphics[width=0.4\textwidth,height=0.1\textwidth]
{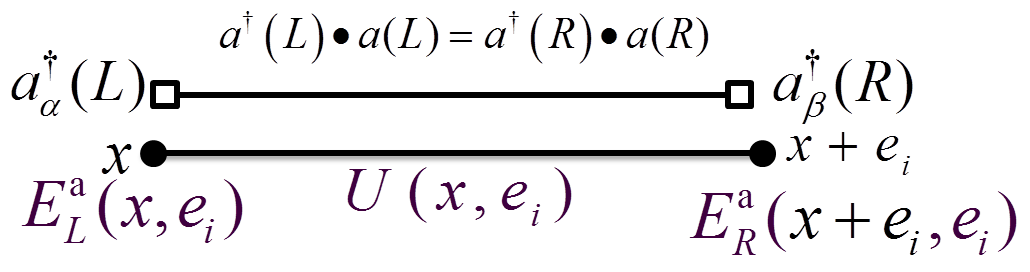}
\end{center}
\caption{Prepotentials on a link} 
\label{su2pp}
\end{figure}
 Using the Schwinger Bosons construction of the angular momentum algebra (\ref{eec}), the 
left and the right electric fields on a link $(x,e_i)$ can be written as: 
\bea
\label{sb} 
\mbox{Left electric fields:} ~~ \quad \quad \quad E_L^{\mathrm a}(x,e_i) &\equiv& 
a^{\dagger}(x,e_i;L)\frac{\sigma^{\mathrm a}}{2}a(x,e_i;L), \\ 
\mbox{Right electric fields:} \quad ~ E_R^{\mathrm a}(x+e_i,e_i) &\equiv & 
a^{\dagger}(x+e_i,e_i;R)\frac{\sigma^{\mathrm a}}{2}a(x+e_i,e_i;R). \nonumber 
\eea
From now on we will suppress the index $(x,e_i)$ with the prepotential operators whenever we consider one single link at a time. 

Using (\ref{sb}), the electric field constraint (\ref{E2=e2}) on any link becomes the following number operator 
constraints in terms of the prepotential operators: 
\bea 
\hat{n}(L) \equiv a^\dagger(L) \cdot a(L) = \hat{n}(R) \equiv a^\dagger(R) \cdot a(R) \equiv \hat{n} 
\label{noc} 
\eea 
In (\ref{noc}), $\hat{n} \equiv \hat{n}(x,e_i)$. 
Note that, this is indeed the most novel feature of prepotential formulation, where the
non-Abelian fluxes can be absorbed locally at a site and the Abelian fluxes spread along the links. 
Both the gauge symmetries together lead to non-local (involving at least a plaquette) Wilson loop 
states. \\
In order to construct the Wilson loop states in terms of prepotentials, it is first necessary to construct link operators on each link in terms of Schwinger Bosons. 
From $SU(2)$ gauge transformations of the link operator in (\ref{eugt}) 
and $SU(2) \otimes U(1)$ gauge transformations properties of 
the Schwinger Bosons, we write the link operator of the form
\bea
\label{su2U}
U^{\alpha}{}_{\beta}=\frac{1}{\sqrt{\hat n+1}}\left(\tilde{a}^{\dagger\alpha}(L) \, a^{\dagger}_\beta(R) 
+ a^\alpha(L)  \, \tilde{a}_\beta(R)\right)\frac{1}{\sqrt{\hat n+1}}, 
\eea
The above link operators and electric field satisfies the same canonical commutation relations (\ref{eec}) and (\ref{ccr}) together with  the property
\bea
UU^\dagger=U^\dagger U=1 ~~\& ~~ \mbox{Det}\,U
=1\eea
The loop operators for a gauge theory are constructed by taking the trace of the path ordered product of link operators around any closed curve. Loop operators acting on strong coupling vacuum creates the loop states of the theory. The novel feature of the prepotential formulation is that, the loop operators around any closed path, when re-expressed in terms of Schwinger Bosons turns out to be direct product of gauge invariant operators at each site. We call those local gauge invariant operators as the local linking operators of the theory and linking states are created by the action of linking operators on strong coupling vacuum. The linking variables together with the Abelian Gauss law constitutes the loop variables of the theory.

\section{Linking Operators, Linking States and  the Diagrammatica}

In this section we explicitly illustrate all possible linking operators and linking states present at each site of a 2 dimensional spatial lattice. We also develop a diagrammatic prescription to illustrate the linking operators and their actions on an arbitrary linking state, which turns out to be extremely useful in the study of the Hamiltonian and its dynamics in later sections.

\begin{figure}[h]
\begin{center}
\includegraphics[width=0.2\textwidth,height=0.2\textwidth]
{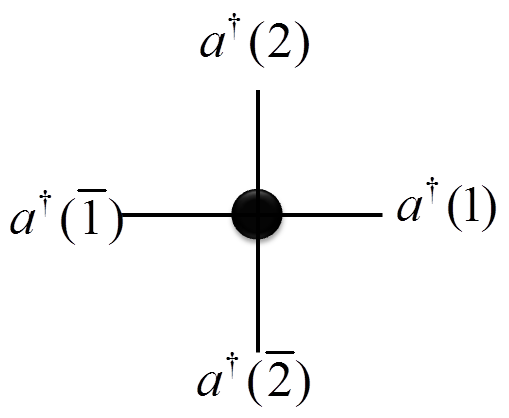}
\end{center}
\caption{A particular site on a 2 dimensional lattice and associated prepotentials} 
\label{dirnot}
\end{figure}

\begin{figure}[h]
\begin{center}
\includegraphics[width=0.2\textwidth,height=0.2\textwidth]
{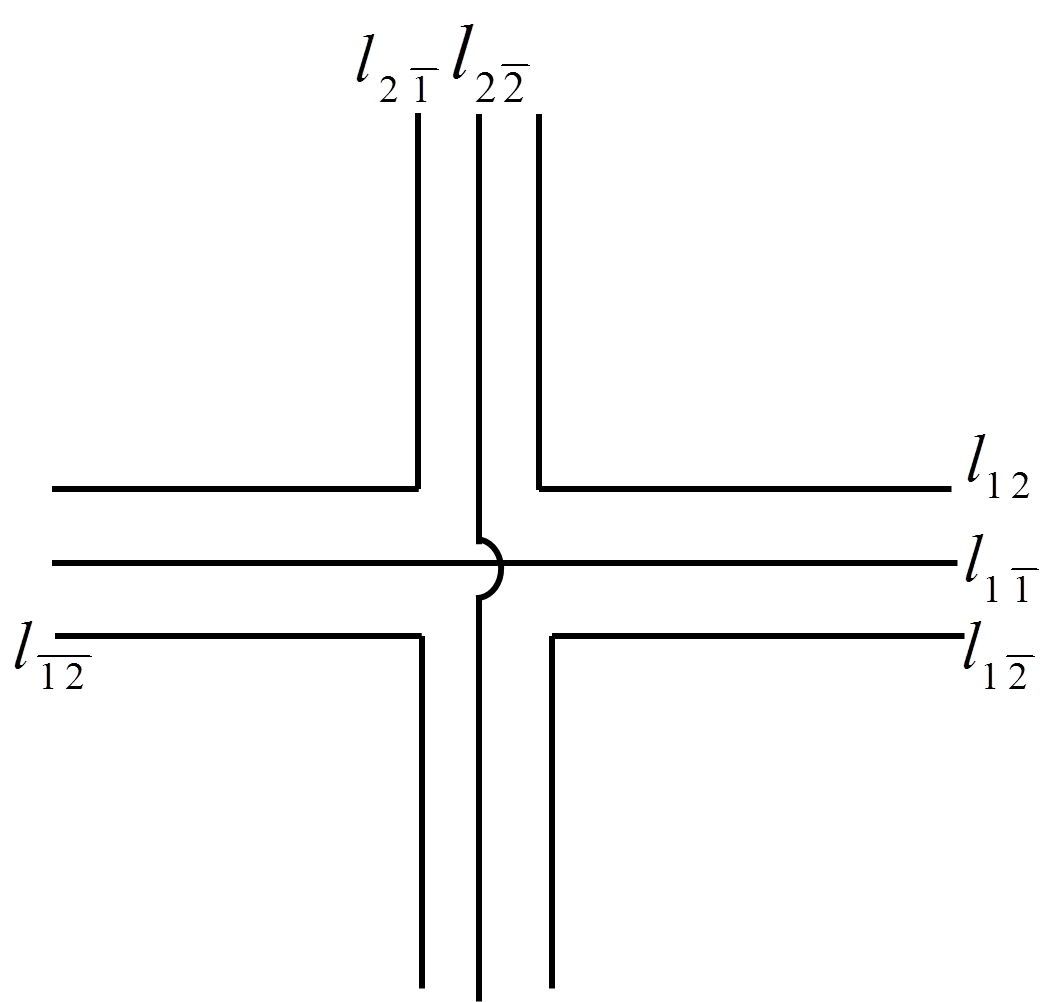}
\end{center}
\caption{SU(2) fluxes: all possible linking at a site of a two dimensional lattice} 
\label{lij_fig}
\end{figure}

We first concentrate at a particular site of a $2$-dimensional spatial lattice, where, $4$ links meet, each link carries its own link operator as given in (\ref{su2U}). There exists four basic local gauge invariant operators (constructed by $U^{\alpha}{}_{\beta}(x,e_i)U^{\beta}{}_{\gamma}(x+e_i,e_j)$ at site $(x+e_i)$) which we  list below:
\bea
\hat{\mathcal O}^{i_+j_+}&\equiv & a^{\dagger}_\beta(i)\frac{1}{\sqrt{\hat n_i +1}}\frac{1}{\sqrt{\hat n_j +1}}\tilde a^{\dagger\beta}(j)= \frac{1}{\sqrt{\hat n_i}}\frac{1}{\sqrt{\hat n_j +1}}a^{\dagger}(i)\cdot\tilde a^{\dagger\beta}(j)\equiv \frac{1}{\sqrt{\hat n_i(\hat n_j+1)}}k_+^{ij} \label{k+} \\
\label{k+-}
\hat{\mathcal O}^{i_+j_-}&\equiv & a^\dagger_\beta(i) \frac{1}{\sqrt{\hat n_i +1}}\frac{1}{\sqrt{\hat n_j +1}} a^\beta(j) = \frac{1}{\sqrt{\hat n_i}} a^\dagger(i)\cdot a(j)\frac{1}{\sqrt{(\hat n_j+2)}}\equiv \frac{1}{\sqrt{\hat n_i}}\kappa^{ij}\frac{1}{\sqrt{(\hat n_j+2)}}\\
\label{k-+}
\hat{\mathcal O}^{j_+i_-}&\equiv & \tilde a_\beta(i) \frac{1}{\sqrt{\hat n_i +1}}\frac{1}{\sqrt{\hat n_j +1}} \tilde a^{\dagger\beta}(j) = \frac{1}{\sqrt{(\hat n_j+1)}} a(i)\cdot a^\dagger(j)\frac{1}{\sqrt{n_i+1}}\equiv \frac{1}{\sqrt{(\hat n_j+1)}}\kappa^{ji}\frac{1}{\sqrt{n_i+1}} ~~~~~\\ 
\label{k-}
\hat{\mathcal O}^{i_-j_-}&\equiv & \tilde a_\beta(i) \frac{1}{\sqrt{\hat n_i +1}}\frac{1}{\sqrt{\hat n_j +1}}  a^{\beta}(j) = \tilde a(i)\cdot a(j)\frac{1}{\sqrt{(\hat n_i+1)(\hat n_j+2)}}\equiv k_-^{ji}\frac{1}{\sqrt{(\hat n_i+1)(\hat n_j+2)}}~~~~~~
\eea              
where, the labels $(i/j)$ associated with prepotential operators actually denote the prepotentials associated with the links along $(i/j)$ directions at that site $x$. For $d=2$, $i,j$ can take values $1,2,\bar 1,\bar 2$ and each direction contains a prepotential doublet $a^\dagger(i)$ as shown in figure \ref{dirnot} .
The maximally commuting gauge invariant set of operators $k_+^{ij}$'s are called linking operators and $k_-^{ij}$'s are their conjugates.
The linking states are constructed by the action of linking operators on strong coupling vacuum. Thus in the prepotential formulation, defining the linking operators locally at each site enables us to define the linking states also locally at each site as,
\bea
|l_{ij}\rangle= \frac{\left(k_+^{ij}\right)^{l_{ij}}}{l_{ij}!}|0\rangle.
\label{lsd}
\eea 
In the prepotential approach, as defined in (\ref{lsd}), the linking states are naturally characterized by the linking quantum numbers  $l_{ij}$, which counts the flux along $i-j$ direction. On a 2d lattice, four links in direction $i$, (with $i=1,2,\bar 1,\bar 2$) meet at  a site, each carrying its own prepotential $a^\dagger(i)$.
Note that $k_+^{ji}=-k_+^{ij}$ by construction given in ((\ref{k+})), makes the loop space in two spatial dimension, to be characterized by six linking numbers $l_{ ij}$, for $i< j$ with the convention that $1<2<\bar 1<\bar 2$. 
 Thus  the most general gauge invariant states at a particular site are characterized by the six liking quantum numbers as follows:
\bea
|l_{12},l_{1\bar 1},l_{1\bar 2},l_{2\bar 1},l_{2\bar 2},l_{\bar 1\bar 2}\rangle \equiv |\{l\}\rangle= \frac{\left(k_+^{12}\right)^{l_{12}}}{l_{12}!}\frac{\left(k_+^{1\bar 1}\right)^{l_{1\bar 1}}}{l_{1\bar 1}!}\frac{\left(k_+^{1\bar 2}\right)^{l_{1\bar 2}}}{l_{1\bar 2}!}\frac{\left(k_+^{2\bar 1}\right)^{l_{2\bar 1}}}{l_{2\bar 1}!}\frac{\left(k_+^{2\bar 2}\right)^{l_{2\bar 2}}}{l_{2\bar 2}!}\frac{\left(k_+^{\bar 1\bar 2}\right)^{l_{\bar 1\bar 2}}}{l_{\bar 1\bar 2}!}|0\rangle \label{lij}
\eea 
 From the definition of the state (\ref{lij}), one can relate the number of prepotential operators at each link to the linking quantum numbers in the following way:
\bea
\label{n1}
n_1 &=& l_{12}+l_{1\bar 1}+l_{1\bar 2}\\
\label{n2}
n_2 &=& l_{2\bar 1}+l_{2\bar 2}+l_{12}\\
\label{n3}
n_{\bar 1} &=& l_{\bar 1\bar 2}+l_{1\bar 1}+l_{2\bar 1}\\
\label{n4}
n_{\bar 2} &=& l_{1\bar 2}+l_{2\bar 2}+l_{\bar 1\bar 2}
\eea
These numbers are basically eigenvalues of the operators $\hat n_i\equiv a^\dagger(i)\cdot a(i)$.
The linking quantum numbers are pictorially represented for a two dimensional lattice in figure \ref{lij_fig}. 

We now illustrate the action of the linking operators defined in (\ref{k+}), (\ref{k+-}) and (\ref{k-}) on the linking states defined in (\ref{lij}). We also prescribe a diagrammatic realization of these actions, which seems to be much more convenient than dealing with long mathematical expressions. The Mathematical expression can be read off from the diagrams by a set of rules given later in this section.

\begin{figure}[h]
\begin{center}
\includegraphics[width=0.2\textwidth,height=0.1\textwidth]
{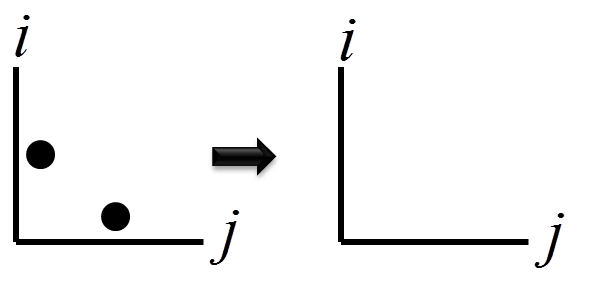}
\end{center}
\caption{The left and right hand side of $=$ denotes the respective sides of the equation (\ref{ac++}) with all the coefficients.} 
\label{ac++fig}
\end{figure}

The basic local gauge invariant operators arising at a particular site as given in  (\ref{k+}-\ref{k-}) are $\hat{\mathcal O}^{i_+j_+}$, $\hat{\mathcal O}^{i_+j_-}$ and $\hat{\mathcal O}^{i_-j_-}$.
The first one acts trivially on the states (\ref{lij}), and increases the flux along $i-j$ direction by one unit. With proper factors in the definition  of the state in (\ref{lij}) as well as the linking operators in (\ref{k+}), the explicit action is obtained as:
\bea
\label{ac++}
\hat{\mathcal O}^{i_+j_+}|\{l\}\rangle &\equiv &  \frac{(l_{ij}+1)}{\sqrt{( n_i+1)( n_j+2)}}|l_{ij}+1\rangle
\eea
Pictorially (\ref{ac++}) is represented in figure \ref{ac++fig}.
Note that, in figure \ref{ac++fig}, the left hand side contain solid dots on solid line. The solid dot denotes the operators acting on a state, more specifically dot on a solid line represents prepotential  creation operator corresponding to that direction acts on a general state. The right hand side of the equation does not contain any dot and represents the state created. Any solid linking line passing through $i-j$ direction at a site, denotes that in the new state the flux along that particular direction has increased by one unit. Note that, in the pictures we are suppressing the symbols for the state for brevity. The coefficients in (\ref{ac++}) are all subsumed in figure \ref{ac++fig}. The algebra towards \ref{ac++} is given in Appendix A.

\begin{figure}[h]
\begin{center}
\includegraphics[width=0.3\textwidth,height=0.15\textwidth]
{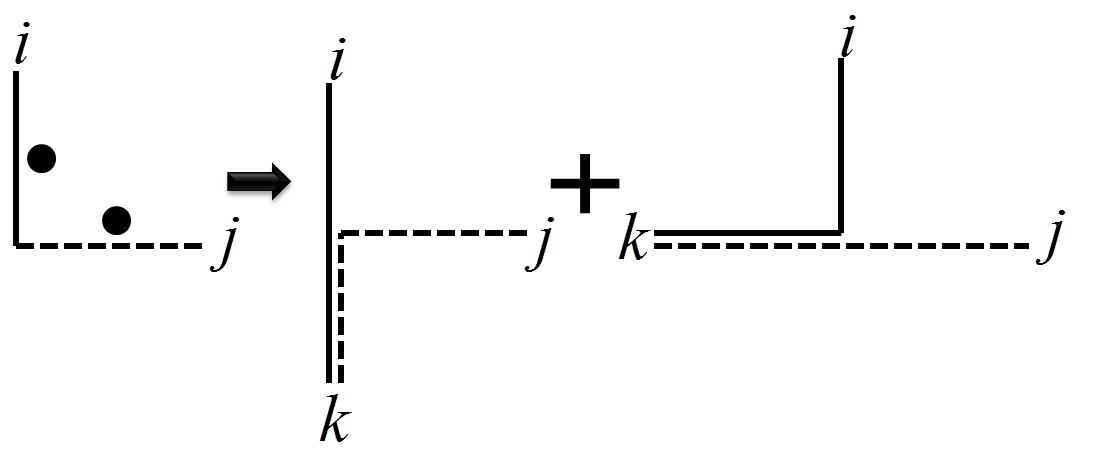}
\end{center}
\caption{The left and right hand side of $=$ denotes the respective sides of the equation (\ref{ac+-}) with all the coefficients.} 
\label{ac+-fig}
\end{figure}

Next we consider the action of (\ref{k+-}) on a general linking state. This action is a bit complicated as one need to use all the commutation relations between different $k_+^{ij}, k_-^{ij}$ and $\kappa^{ij}$ to move the annihilation operator towards the right. However, after the algebraic simplification (as shown in Appendix A) the action of the operators defined in (\ref{k+-}), on any arbitrary linking state is obtained as:
\bea
\label{ac+-}
\hat{\mathcal O}^{i_+j_-}|\{l\}\rangle \equiv   \frac{1}{\sqrt{(n_i+1)(n_j+2)}}\sum_{k\ne i,j} (-1)^{S_{ik}}(l_{ik}+1) |l_{jk}-1,l_{{ik}}+1\rangle
\eea
where, in any of the $l_{ij}$'s in the above equation (and also in any equation throughout the paper), the indices are by-default considered to be rearranged in such a way, that  the first index is always less than the second one in accordance with the ordering convention $
1<2<\bar 1<\bar 2$. 
The factor $S_{ik}$ is calculated as,
\bea
S_{ik}=1~~\mbox{if }i>k~~\&~~S_{ik}=0~~\mbox{if }i<k. 
\label{sik}
\eea
We represent the action of the gauge invariant operator in (\ref{ac+-}) pictorially in figure \ref{ac+-fig}. In the left hand side of figure \ref{ac+-fig}, the solid dot on solid line denotes prepotential creation operator along that direction and solid dot on dashed line denotes annihilation operator along that direction acts on the state. In right hand side, the dashed line represents that the corresponding solid line in the state is removed if it was already present in the state and it is zero if there were none already present.
Note that as given in (\ref{ac+-}), each term comes with a particular coefficient which we absorb in the diagrams itself. This is possible by providing with a set of rules (similar to the Feynman rules) for associating each diagram with the coefficient. Having exhausted with all possible linking actions on a general linking states, we will state all of the rules at the end of this section. These new lattice Feynman rules will enable us to do any loop computation diagrammatically.

\begin{figure}[h]
\begin{center}
\includegraphics[width=0.5\textwidth,height=0.15\textwidth]
{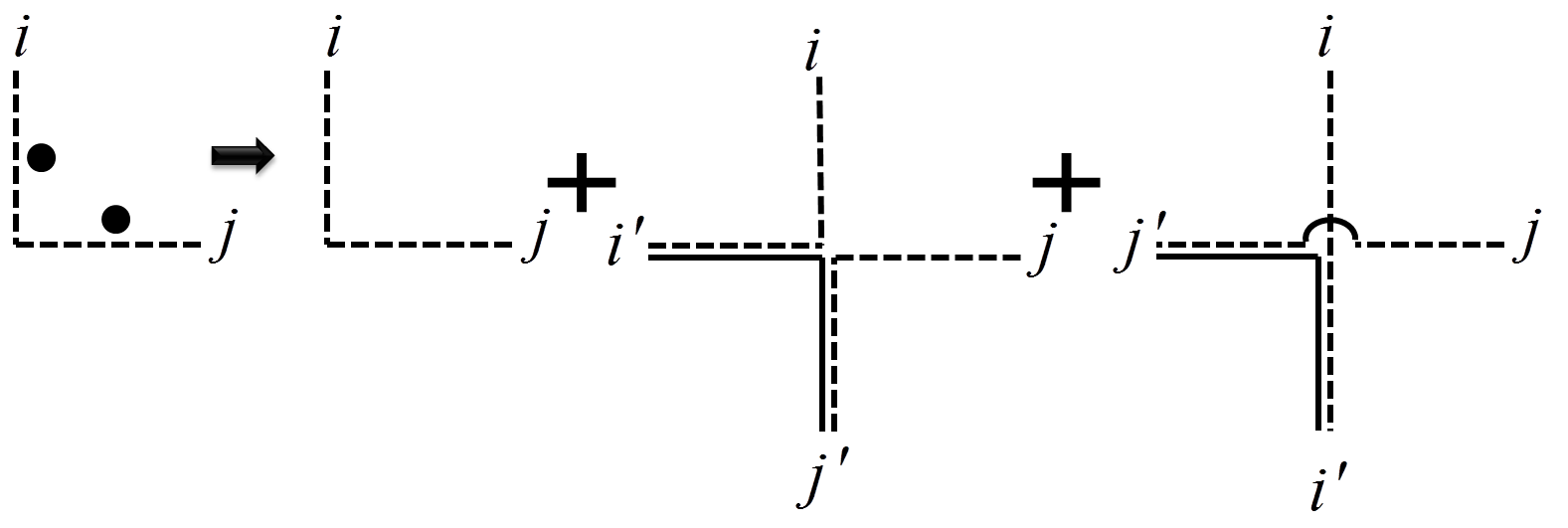}
\end{center}
\caption{The left and right hand side of $=$ denotes the respective sides of the equation (\ref{ac--}) with all the coefficients. Note that the usual vertex symbol denotes the unusual coefficient for the first term of the decomposition.} 
\label{ac--fig}
\end{figure}

Let us next consider the remaining local gauge invariant operator  $\hat{\mathcal O}^{i_-j_-}$ and its action on a general linking state. This action is the most complicated one to calculate as both the annihilation operators are needed to move to right by using the commutation relations. A long calculation given in Appendix A finally yields the following action,
\bea
\label{ac--}
\hat{\mathcal O}^{i_-j_-}|\{l\}\rangle &=& \frac{1}{\sqrt{(n_i+1)(n_j+2)}}\Bigg[ (n_i+n_j-l_{ij}+1)|l_{ij}-1\rangle \nonumber \\ &&+  \sum_{i', j'\{\ne i,j\}}(l_{i'j'}+1)(-1)^{S_{i'j'}}| l_{ii'}-1,l_{jj'}-1,l_{i' j'}+1\rangle  \Bigg]~~~~~~~
\eea
To realize the action better, one can find its pictorial representation as in figure \ref{ac--fig}. The first term in the right hand side of (\ref{ac--}), is the simplest one and is given by the first diagram in the right hand side of the figure \ref{ac--fig}. However, the terms within the summation in (\ref{ac--}), gives rise to two terms for two dimensional spatial lattice as shown in figure \ref{ac--fig}.

\begin{figure}[h]
\begin{center}
\includegraphics[width=0.7\textwidth,height=0.3\textwidth]
{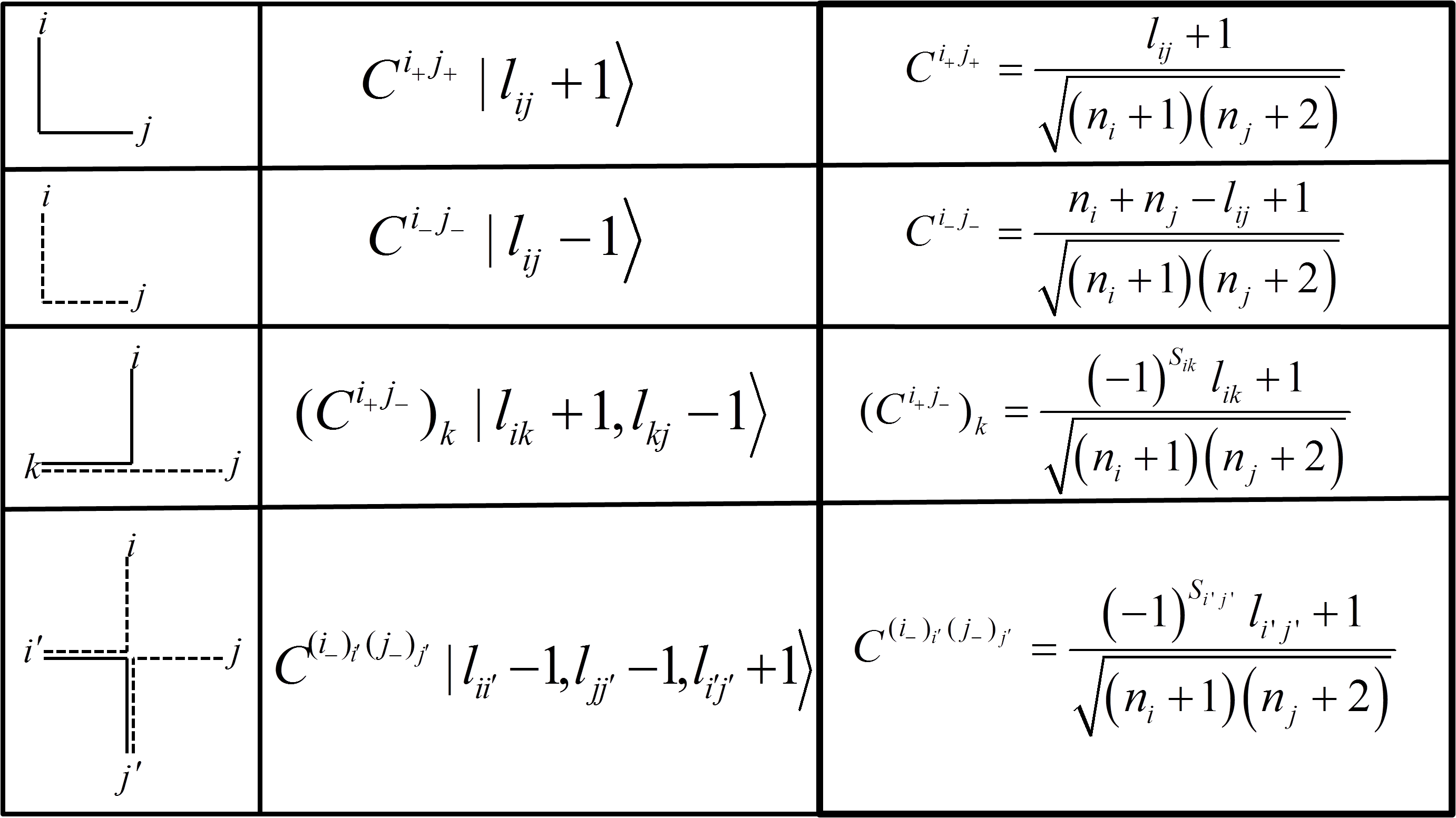}
\end{center}
\caption{The coefficients are explicitly given in the last column.} 
\label{rule}
\end{figure}

The actions of local gauge invariant operators (constructed out of prepotential operators) on the linking states characterized by linking quantum numbers in (\ref{lij})  are obtained in (\ref{ac++},\ref{ac+-},\ref{ac--}), and pictorially represented in figures \ref{ac++fig}, \ref{ac+-fig}, \ref{ac--fig}. Note that, the pictorial representation of the states contain the particular coefficients appearing before the states in any of (\ref{ac++},\ref{ac+-},\ref{ac--}) along with the states produced characterized by the linking numbers. Hereby we prescribe a set of rules to read off the coefficient as well as the state by just looking at a particular diagram! Hence a particular diagram would correspond to a state characterized by linking numbers with a coefficient sitting in front of it as shown in the table in \ref{rule}.

Now, from the coefficients given above and the diagrams in \ref{rule}, we can spell out the `lattice Feynman rules' as follows:
\begin{itemize}
\item Any diagram with net flux increasing or decreasing along $i-j$ direction (or increasing along $i$  and decreasing along $j$ directions together) contribute a factor of $\frac{1}{\sqrt{(n_i+1)(n_j+2)}}$, where $n_i,n_j$ counts the flux of the state on which the loop operator has acted.
\item Each solid line crossing the site from direction $i-j$ will contribute a factor of $l_{ij}+1$. 
\item Each dotted line crossing the site from direction $i-j$, without having any overlap with any solid line on any of its arm, will contribute a factor of $(n_i+n_j-l_{ij}+1)$.
\item Each solid flux line along $i-k$ direction with the link at $k$ direction, having overlap with a dotted link along $k-j$ direction will contribute a factor of $(-1)^{S_{ik}}$ defined in (\ref{sik}).
\item Each solid flux line along $i-j$ direction with the link at $i$ direction, having overlap with a dotted link along $i-i'$ direction and the link at $j$ direction, having overlap with a dotted link along $j-j'$ direction will contribute a factor of $(-1)^{S_{ij}}$ defined in (\ref{sik}), where $i'<j'$.
\end{itemize}
To make the above diagrammatic rules more clear, we tabulate all possible loop configurations that can occur at each of the four vertices (namely $a,b,c,d$) of a plaquette, by the action of local gauge invariant operators at the same in the following table. Note that, these loop configurations are obtained in the dynamics of loops under the magnetic Hamiltonian, as discussed in detail in the next section.
\begin{center}
\begin{tabular}{|c|c|c|c|}
\hline
vertex & coefficient & vertex & coefficient  \\
\hline
&&&\\
d1: \includegraphics[width=5mm, height=5mm]{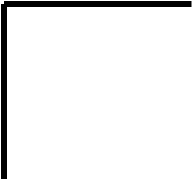} & $C^{1_+\bar 2_+}= \frac{l_{1\bar 2}+1}{\sqrt{(n_1+1)(n_{\bar 2}+2)}}$ & a1: \includegraphics[width=5mm, height=5mm]{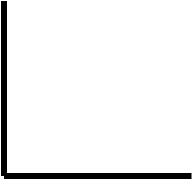} & $C^{1_+ 2_+}= \frac{l_{12}+1}{\sqrt{(n_1+1)(n_{ 2}+2)}}$\\
\hline
&&&\\
d2: \includegraphics[width=5mm, height=5mm]{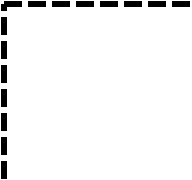} & $C^{1_-\bar 2_-}=  \frac{(n_1+n_{\bar 2}-l_{1\bar 2}+1)}{\sqrt{(n_1+1)(n_{\bar 2}+2)}}$
&a2: \includegraphics[width=5mm, height=5mm]{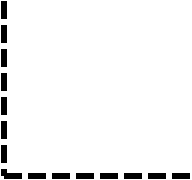} & $C^{1_-2_-}=  \frac{(n_1+n_{2}-l_{12}+1)}{\sqrt{(n_1+1)(n_{2}+2)}}$ \\
\hline
&&&\\
d3: \includegraphics[width=8mm, height=5mm]{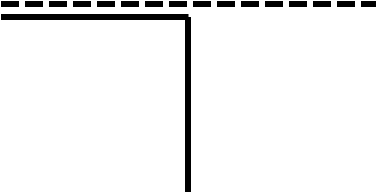} & $\left(C^{\bar 2_+1_-}\right)_{\bar 1}=  -\frac{l_{\bar 1\bar 2}+1}{\sqrt{(n_{\bar 2}+1)(n_{1}+2)}}$ &
a3: \includegraphics[width=8mm, height=5mm]{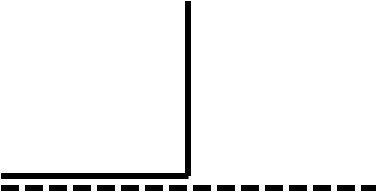} & $\left(C^{2_+1_-}\right)_{\bar 1}=  \frac{l_{2\bar 1}+1}{\sqrt{(n_{2}+1)(n_{1}+2)}}$ \\
\hline
&&&\\
d4: \includegraphics[width=5mm, height=10mm]{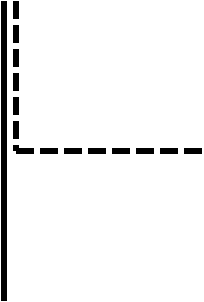}  & $\left(C^{\bar 2_+1_-}\right)_{2}=  -\frac{l_{2\bar 2}+1}{\sqrt{(n_{\bar 2}+1)(n_{1}+2)}}$ &
a4: \includegraphics[width=5mm, height=10mm]{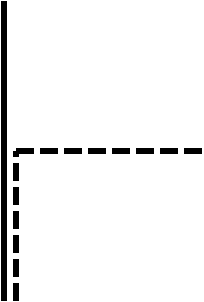}  & $\left(C^{2_+1_-}\right)_{\bar 2}=  \frac{l_{2\bar 2}+1}{\sqrt{(n_{2}+1)(n_{1}+2)}}$\\
\hline
&&&\\
d5:  \includegraphics[width=8mm, height=5mm]{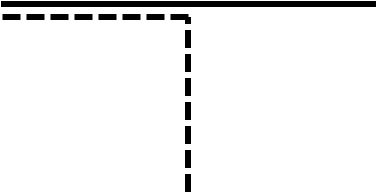} & $\left(C^{1_+\bar 2_-}\right)_{\bar 1}=  \frac{l_{1\bar 1}+1}{\sqrt{(n_{1}+1)(n_{\bar 2}+2)}}$ &
a5:  \includegraphics[width=8mm, height=5mm]{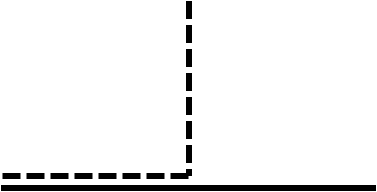} & $\left(C^{1_+2_-}\right)_{\bar 1}=  \frac{l_{1\bar 1}+1}{\sqrt{(n_{1}+1)(n_{ 2}+2)}}$ \\
\hline
&&&\\
d6:  \includegraphics[width=5mm, height=10mm]{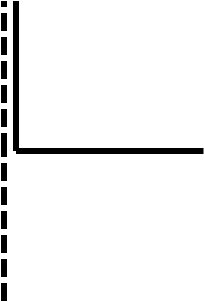} & $\left(C^{1_+\bar 2_-}\right)_{2}=  \frac{l_{12}+1}{\sqrt{(n_{1}+1)(n_{\bar 2}+2)}}$ &
a6:  \includegraphics[width=5mm, height=10mm]{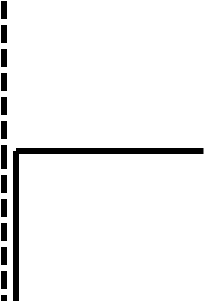} & $\left(C^{1_+2_-}\right)_{\bar 2}=  \frac{l_{1\bar 2}+1}{\sqrt{(n_{1}+1)(n_{ 2}+2)}}$ \\
\hline
d7:\includegraphics[width=8mm, height=8mm]{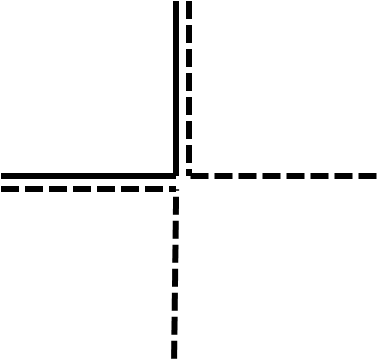} & $C^{(1_-)_2(\bar 2_-)_{\bar 1}}= \frac{l_{2\bar 1}+1}{\sqrt{(n_1+1)(n_{\bar 2}+2)}}$ &
a7:\includegraphics[width=8mm, height=8mm]{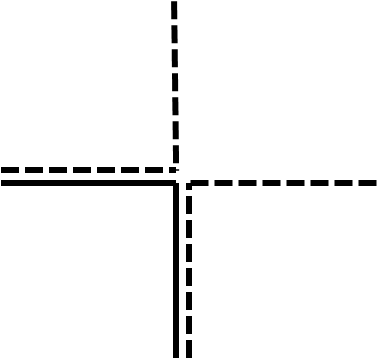} & $C^{(1_-)_{\bar 2}(2_-)_{\bar 1}}= -\frac{l_{\bar 1\bar 2}+1}{\sqrt{(n_1+1)(n_{2}+2)}}$ \\
\hline
d8:\includegraphics[width=8mm, height=8mm]{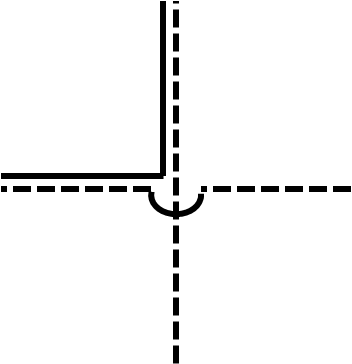} & $C^{(1_-)_{\bar 1}(\bar 2_-)_{2}}= -\frac{l_{2\bar 1}+1}{\sqrt{(n_1+1)(n_{\bar 2}+2)}}$ &
a8:\includegraphics[width=8mm, height=8mm]{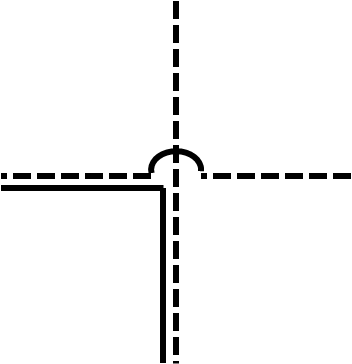} & $C^{(1_-)_{\bar 1}(2_-)_{\bar 2}}= \frac{l_{\bar 1\bar 2}+1}{\sqrt{(n_1+1)(n_{2}+2)}}$ \\
\hline
\hline
&&&\\
b1:\includegraphics[width=5mm, height=5mm]{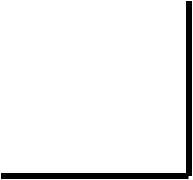} & $C^{\bar 1_+2_+}= \frac{l_{\bar 1 2}+1}{\sqrt{(n_{\bar 1}+1)(n_{2}+2)}}$ & 
c1: \includegraphics[width=5mm, height=5mm]{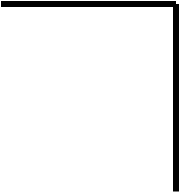}& $C^{\bar 1_+ \bar 2_+}= \frac{l_{\bar 1\bar 2}+1}{\sqrt{(n_{\bar 1}+1)(n_{ \bar 2}+2)}}$\\
\hline
&&&\\
b2: \includegraphics[width=5mm, height=5mm]{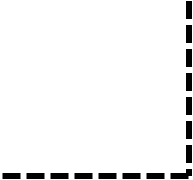}& $C^{\bar 1_-2_-}=  \frac{(n_{\bar 1}+n_{2}-l_{\bar 1 2}+1)}{\sqrt{(n_{\bar 1}+1)(n_{ 2}+2)}}$
&c2:\includegraphics[width=5mm, height=5mm]{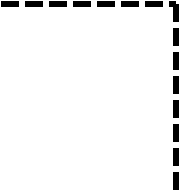} & $C^{\bar 1_-\bar 2_-}=  \frac{(n_{\bar 1}+n_{\bar 2}-l_{\bar 1\bar 2}+1)}{\sqrt{(n_{\bar 1}+1)(n_{\bar 2}+2)}}$ \\
\hline
&&&\\
b3:\includegraphics[width=8mm, height=5mm]{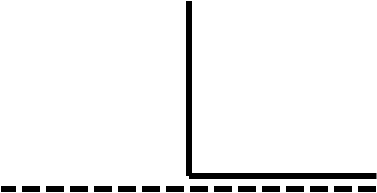} & $\left(C^{2_+\bar 1_-}\right)_{1}=  -\frac{l_{12}+1}{\sqrt{(n_{ 2}+1)(n_{\bar 1}+2)}}$ &
c3:\includegraphics[width=8mm, height=5mm]{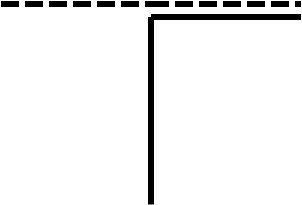} & $\left(C^{\bar 2_+\bar 1_-}\right)_{1}=  -\frac{l_{1\bar 2}+1}{\sqrt{(n_{\bar 2}+1)(n_{\bar 1}+2)}}$ \\
\hline
&&&\\
b4:\includegraphics[width=5mm, height=8mm]{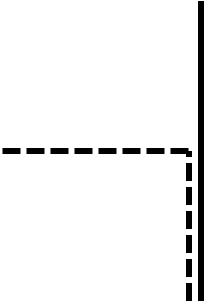} & $\left(C^{2_+\bar 1_-}\right)_{\bar 2}=  \frac{l_{2\bar 2}+1}{\sqrt{(n_{ 2}+1)(n_{\bar 1}+2)}}$ &
c4:\includegraphics[width=5mm, height=8mm]{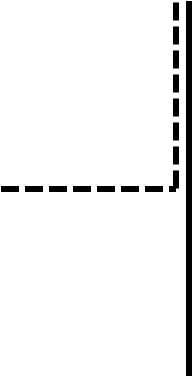} & $\left(C^{\bar 2_+\bar 1_-}\right)_{2}=  -\frac{l_{2\bar 2}+1}{\sqrt{(n_{\bar 2}+1)(n_{\bar 1}+2)}}$\\
\hline
&&&\\
b5:\includegraphics[width=8mm, height=5mm]{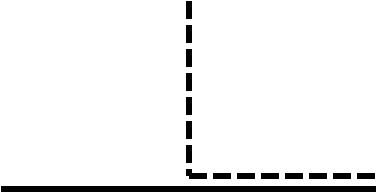} & $\left(C^{\bar 1_+ 2_-}\right)_{1}=  -\frac{l_{1\bar 1}+1}{\sqrt{(n_{\bar 1}+1)(n_{ 2}+2)}}$ &
c5:\includegraphics[width=8mm, height=5mm]{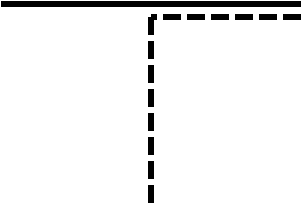} & $\left(C^{\bar 1_+\bar 2_-}\right)_{1}= - \frac{l_{1\bar 1}+1}{\sqrt{(n_{\bar 1}+1)(n_{ \bar 2}+2)}}$ \\
\hline
&&&\\
b6:\includegraphics[width=5mm, height=8mm]{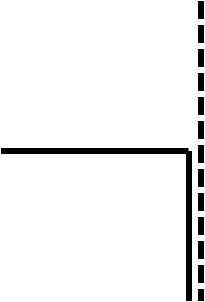} & $\left(C^{\bar 1_+ 2_-}\right)_{\bar 2}=  \frac{l_{\bar 1\bar 2}+1}{\sqrt{(n_{\bar 1}+1)(n_{2}+2)}}$ &
c6:\includegraphics[width=5mm, height=8mm]{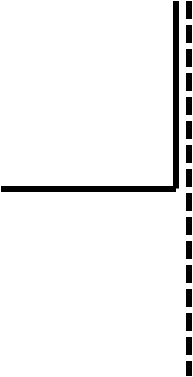} & $\left(C^{\bar 1_+\bar 2_-}\right)_{2}=  -\frac{l_{2\bar 1}+1}{\sqrt{(n_{\bar 1}+1)(n_{\bar  2}+2)}}$ \\
\hline
&&&\\
b7:\includegraphics[width=8mm, height=8mm]{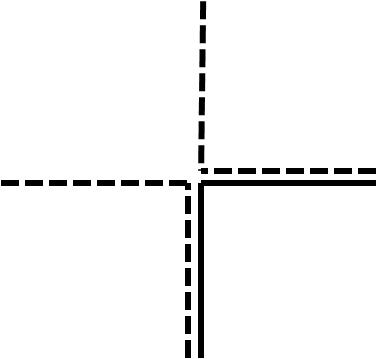} & $C^{(2_-)_{1}(\bar 1_-)_{\bar 2}}= \frac{l_{1\bar 2}+1}{\sqrt{(n_{\bar 1}+1)(n_{ 2}+2)}}$ &
c7:\includegraphics[width=8mm, height=8mm]{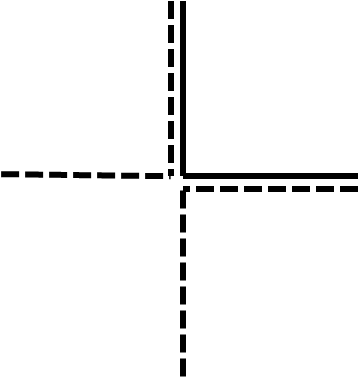} & $C^{(\bar 1_-)_{2}(\bar 2_-)_{1}}= -\frac{l_{12}+1}{\sqrt{(n_{\bar 1}+1)(n_{\bar 2}+2)}}$ \\
\hline
&&&\\
b8:\includegraphics[width=8mm, height=8mm]{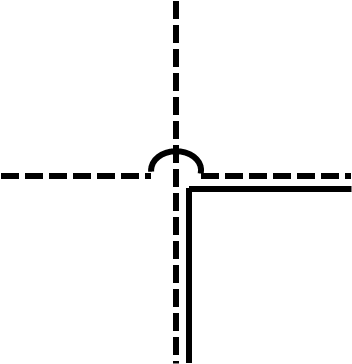} & $C^{(2_-)_{\bar 2}(\bar 1_-)_{1}}= -\frac{l_{1\bar 2}+1}{\sqrt{(n_{\bar 1}+1)(n_{ 2}+2)}}$ &
c8:\includegraphics[width=8mm, height=8mm]{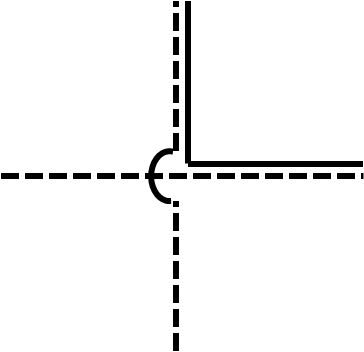} & $C^{(\bar 1_-)_{1}(\bar 2_-)_{2}}= \frac{l_{12}+1}{\sqrt{(n_{\bar 1}+1)(n_{\bar 2}+2)}}$ \\
\hline
\end{tabular}
\end{center}

At this point, we discuss the overcompleteness in the loop basis characterized by linking numbers in the next subsection.

\subsection{Physical Degrees of Freedom}

\begin{figure}[h]
\begin{center}
\includegraphics[width=0.7\textwidth,height=0.15\textwidth]
{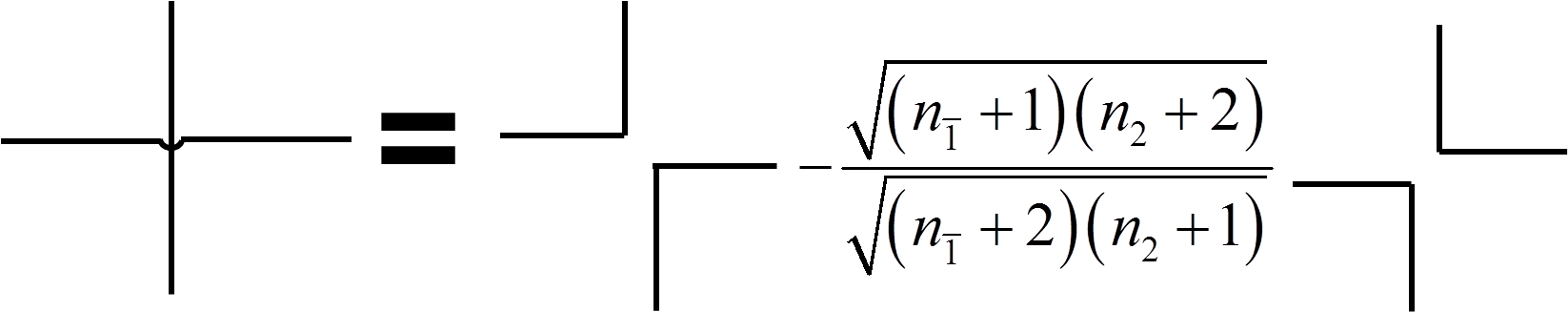}
\end{center}
\caption{Pictorial representation of Mandelstam constraint in terms of prepotentials as given in (\ref{mcor}).} 
\label{k+const}
\end{figure}

\begin{figure}[h]
\begin{center}
\includegraphics[width=0.8\textwidth,height=0.3\textwidth]
{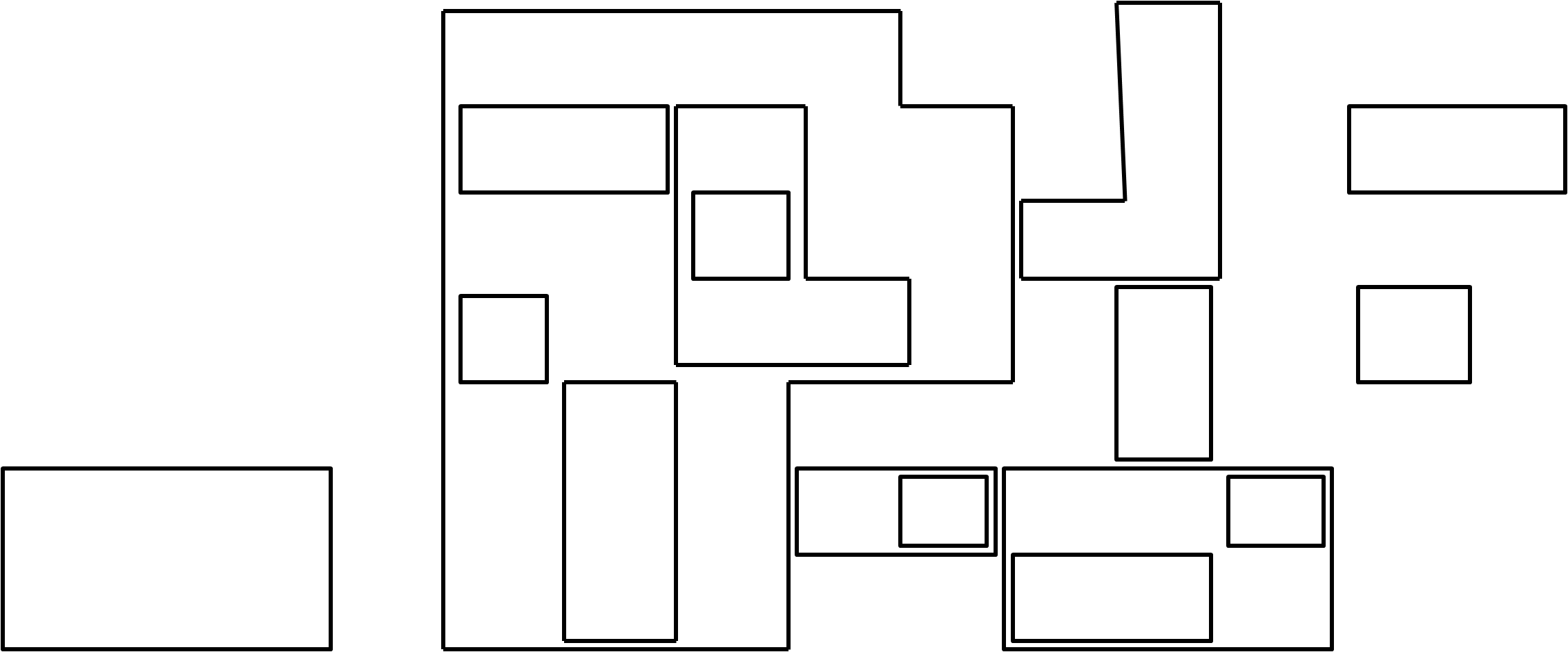}
\end{center}
\caption{Physical Loops: Nested and overlapped but non-intersecting ones } 
\label{physloop}
\end{figure}

We have already discussed that we can describe the local linking states on a 2d lattice,  by a set of six linking numbers defined locally at each site. These set of linking variables form an over-complete basis of the theory as the physical degrees of freedom for SU(2) gauge theory on 2+1 dimensional lattice is only $3$ per lattice site. Hence, there must be three constraints at each lattice site among the linking number variables, to obtain the exact physical degrees of freedom of the theory. Among these three constraints, two are the number operator constraint arising because of the fact that, $E_L^2=E_R^2$ at each site (as given in (\ref{E2=e2})) and is realized in terms of prepotentials in (\ref{noc}). On two spatial dimensions this constraint ($U(1)$ constraint) reads as 
\bea
\label{2dnoc}
n_1(x)=n_{\bar 1}(x+e_1) ~~\& ~~ n_2(x)=n_{\bar 2}(x+e_2)
\eea
where, $n_1,n_2,n_{\bar 1}$ and $n_{\bar 2}$ are defined in (\ref{n1},\ref{n2},\ref{n3},\ref{n4}), and $e_1$ and $e_2$ are unit vectors (in lattice units) along the two directions. In terms of linking numbers, the two number operator constraint reads as:
\bea
&&l_{12}(x)+l_{1\bar 1}(x)+l_{1\bar 2}(x)= l_{1\bar 1}(x+e_1)+l_{2\bar 1}(x+e_1)+l_{\bar 1\bar 2}(x+e_1)\nonumber \\ & ~\& ~& l_{12}(x)+l_{2\bar 1}(x)+l_{2\bar 2}(x)=l_{1\bar 2}(x+e_2)+l_{2\bar 2}(x+e_2)+l_{\bar 1\bar 2}(x+e_2)\label{u1l}
\eea
The other constraint in $2+1$ dimension is the Mandelstam constraint which in Prepotential formulation, at a particular site of a $2$-d lattice reads as the operator relation:
\bea
\label{mcor}
k_+^{1\bar 1}k_+^{2\bar 2}=k_+^{1\bar 2}k_+^{2\bar 1}- k_+^{12}k_+^{\bar 1\bar 2}
\eea
Using the definitions (\ref{ac++}) and (\ref{k+}), we can write (\ref{mcor}) as:
\bea
&& \sqrt{( n_1+1)( n_{\bar 1}+2)}\hat{\mathcal O}^{1_+\bar 1_+}\sqrt{( n_2+1)(n_{\bar 2}+2)}\hat{\mathcal O}^{2_+\bar 2_+}=\nonumber \\ &&\sqrt{( n_1+1)(n_{\bar 2}+2)}\hat{\mathcal O}^{1_+\bar 2_+} \sqrt{(n_2+1)(n_{\bar 1}+2)}\hat{\mathcal O}^{2_+\bar 1_+}- \sqrt{(n_1+1)(n_2+2)}\hat{\mathcal O}^{1_+2_+}\sqrt{( n_{\bar 1}+1)(n_{\bar 2}+2)}\hat{\mathcal O}^{\bar 1_+\bar 2_+}\nonumber \\
& \Rightarrow & \hat{\mathcal O}^{1_+\bar 1_+}\hat{\mathcal O}^{2_+\bar 2_+}=\left[\hat{\mathcal O}^{1_+\bar 2_+} \hat{\mathcal O}^{2_+\bar 1_+}-\sqrt{ \frac{( n_{\bar 1}+1)(n_{2}+2)}{(n_2+1)(n_{\bar 1}+2)}}\hat{\mathcal O}^{1_+2_+}\hat{\mathcal O}^{\bar 1_+\bar 2_+}\right]
\eea
The Mandelstam constraint is pictorially represented in figure \ref{k+const},
from which, we clearly find that, the linking states with two vertical and horizontal flux lines crossing each other a particular lattice site are actually not  independent  states but are a combination of two different states where the flux lines touches each other at that site itself. 
Another useful way of solving the Mandelstam constraints is to note that any local state generated by the combination $k_+^{1\bar 1}k_+^{2\bar 2}$ can be replaced by the right hand side of (\ref{mcor}). That is in terms of linking numbers without any loss of generality, this amounts to choosing $l_{1\bar 1}$ and $l_{2\bar 2}$ linking numbers at any site, such that,
\bea
\label{mcsol}
l_{1\bar 1}(x)\cdot l_{2\bar 2}(x)=0
\eea
 Mandelstam constraint in terms of linking variables is given in (\ref{mcsol}).
The U(1) constraint (\ref{u1l}) and (\ref{mcsol}) define our physical space completely. 
More specifically, the Abelian U(1) constraint (\ref{u1l}) implies that the physical states are closed electric flux loops while constraint (\ref{mcsol}) implies that these flux loops cannot intersect at any site while they can overlap over lines. Hence, our physical states are made of nested electric flux loops which can overlap over portions but can never intersect. An example set of physically allowed loops are given in figure \ref{physloop}.

From our construction of physical states of gauge theory, we have a norm on the states which is not trivial and indeed our choice of basis states are not even orthogonal to each other. This norm is explicitly spelled out in Appendix B.

\section{Loop States and Fusion Operators}

In this section we discuss enumeration of all physical loop states on the entire lattice. Naively these nested loops can be of arbitrary size and shape, therefore their descriptions are non-local as well. We will show by defining the Fusion operators, the description does become local and complete. The key idea follows from the fact that on a single plaquette, any arbitrary number of electric flux plaquette loops are allowed in the physical space. Larger loops can be formed by suitable fusion of such basic plaquette loops, where the newly invented fusion operators play their roles.

\begin{figure}[h]
\begin{center}
\includegraphics[width=0.5\textwidth,height=0.45\textwidth]
{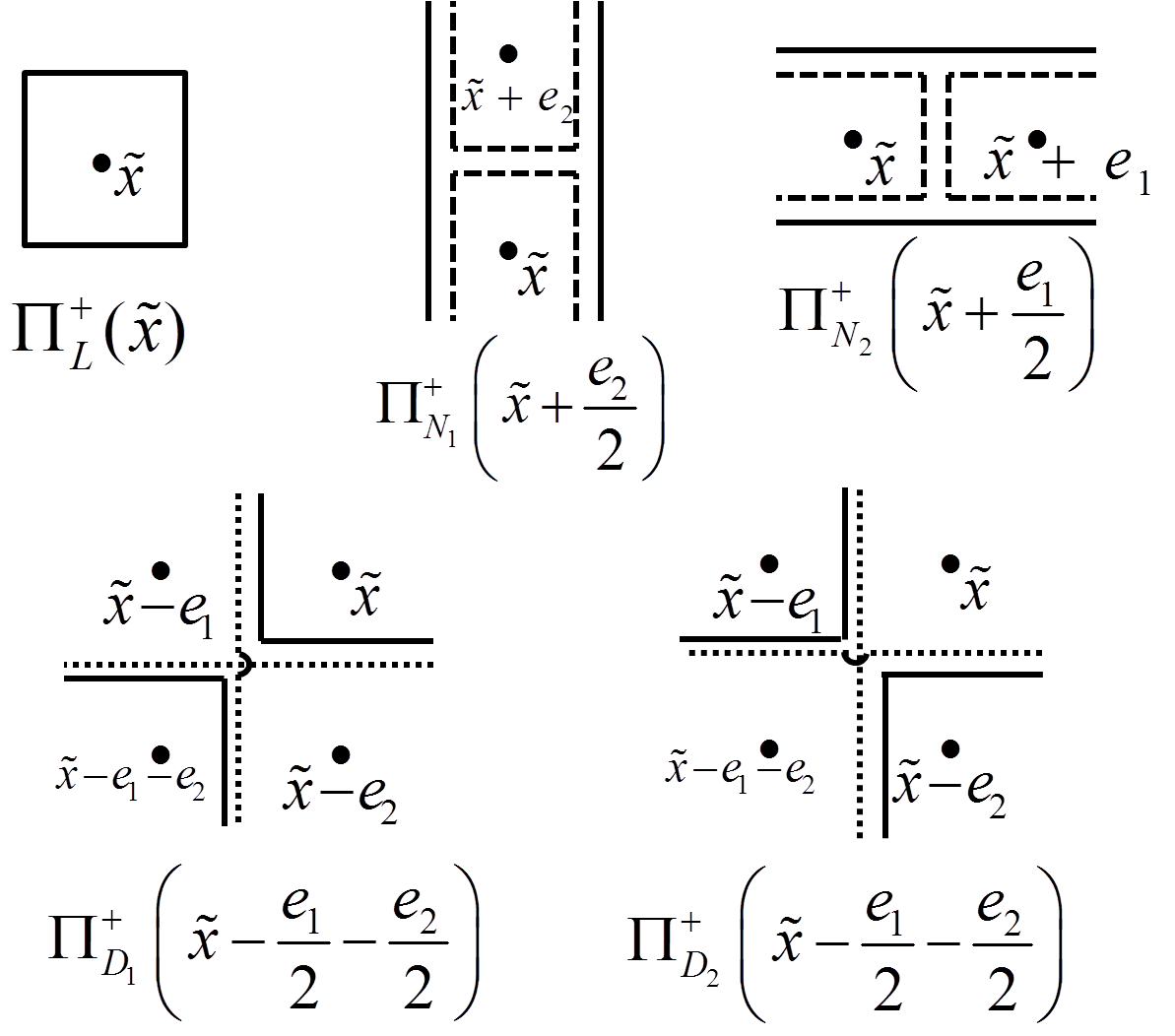}
\end{center}
\caption{Variables defined locally at dual site $(\tilde x)$, on a 2-dimensional lattice spanned by basis vectors $e_1$ along $+X$ axis, and $e_2$ along $+Y$ axis.} 
\label{lnq}
\end{figure}

The simplest way of explaining this construction is by working with diagrammatic technique as given in  figure \ref{lnq}.
In each of these diagrams in figure \ref{lnq}, there is an explicit meaning in terms of the linking operators and the corresponding linking states. To illustrate that clearly, let us understand the following facts:
\begin{itemize}
\item The basic plaquette, the first diagram (a) in figure \ref{lnq}, is the basic electric flux palquette loop and this can be constructed by the action of four linking operators $k_+^{ij}$ at the four vertices around the plaquette on the strong coupling vacuum $|0\rangle$, which we denote as the creation operator $\Pi^+_L(\tilde x)$ acting on $|0\rangle$ and the inverse action, i.e annihilation of a plaquette loop by $\Pi^-_L(\tilde x)$ . $L(\tilde x)$ defines number of such plaquette loops at the dual sites $\tilde x$ of the lattice.
\item Then we define the fusion operators $\Pi^{\pm}_{N_1},\Pi^{\pm}_{N_2},\Pi^{\pm}_{D_1},\Pi^{\pm}_{D_2}$ and the corresponding numbers $N_1,N_2,D_1,D_2$ which construct larger loops by combining neighbouring smaller ones. These fusion variables can be thought of as some operators which either merges two smaller loops to a bigger one, or annihilates the state if no such neighbouring loops are present. In explicit operator form, every solid line in these fusion operators is the $k_+^{ij}$ type of inking operator, while the dashed line is its pseudo-inverse in the sense that, when there is some nonzero flux (denoted by nonzero $l_{ij}$) or solid line already present, the dashed line decreases that by one unit and if none were present, it annihilates.
\item To realize the action of the fusion operators in figure \ref{lnq}, let us consider the following examples:
\begin{itemize}
\item
If  there exists a loop state with $L(\tilde x)=1, L(\tilde x+ e_1)=1$, then there can exist another loop state with  $L(\tilde x)=1, L(\tilde x+ e_1)=1,N_2(\tilde x+\frac{e_1}{2})=1$, which is basically a rectangular loop with horizontal length of two lattice units as shown in figure \ref{eg1}.
\begin{figure}[h]
\begin{center}
\includegraphics[width=0.5\textwidth,height=0.2\textwidth]
{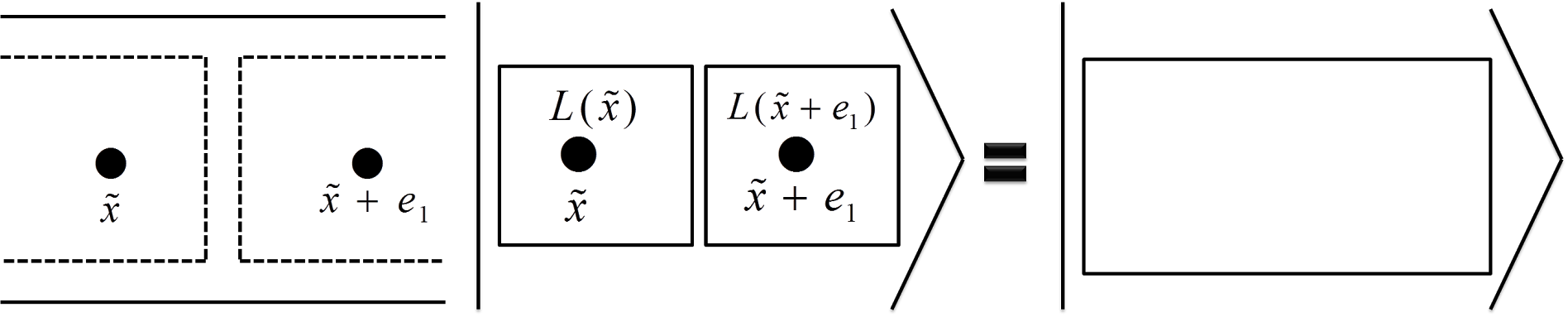}
\end{center}
\caption{ $\Pi^+_{N_2}(\tilde x+\frac{e_1}{2})|L(\tilde x)=1,L(\tilde x+e_1)=1\rangle= |L(\tilde x)=1,L(\tilde x+e_1)=1,{N_2}(\tilde x+\frac{e_1}{2})=1\rangle$} 
\label{eg1}
\end{figure}
 Here, the second state can be thought as created by the fusion operator $\Pi^+_{N_2}(\tilde x+\frac{e_1}{2})$ on the first state. However, applying fusion operator once again would annihilate the state implying no state to exist with $L(\tilde x)=1, L(\tilde x+ e_1)=1,N_2(\tilde x+\frac{e_1}{2})=2$.
\item Similarly the fusion  operator $\Pi^+_{N_1}(\tilde x-\frac{e_2}{2})$ combines vertical neighbouring plaquettes if they are present.
\item The inverse action. i.e decoupling a bigger loop to two smaller loops with an overlap along a vertical or horizontal link is performed by the fusion operators  $\Pi^-_{N_2}(\tilde x+\frac{e_1}{2})$ and $\Pi^-_{N_1}(\tilde x-\frac{e_2}{2})$ respectively.
\item The other two fusion operators $\Pi^\pm_{D_{1(2)}}(\tilde x-\frac{e_1}{2}-\frac{e_2}{2})$ combine the diagonal ones as shown in figure \ref{eg3}. Note that  the individual $\Pi^\pm_{D_{1(2)}}(\tilde x-\frac{e_1}{2}-\frac{e_2}{2})$ operators contain intersecting horizontal and vertical flux lines which is not a part of physical loop space. Hence these particular fusion operators should always come in a certain combination (like the $\Pi^+_{D_1}\Pi^-_{D_2}$) with other fusion variables such that, there exists no intersecting flux lines for the final loop state produced. 
\begin{figure}[h]
\begin{center}
\includegraphics[width=0.6\textwidth,height=0.15\textwidth]
{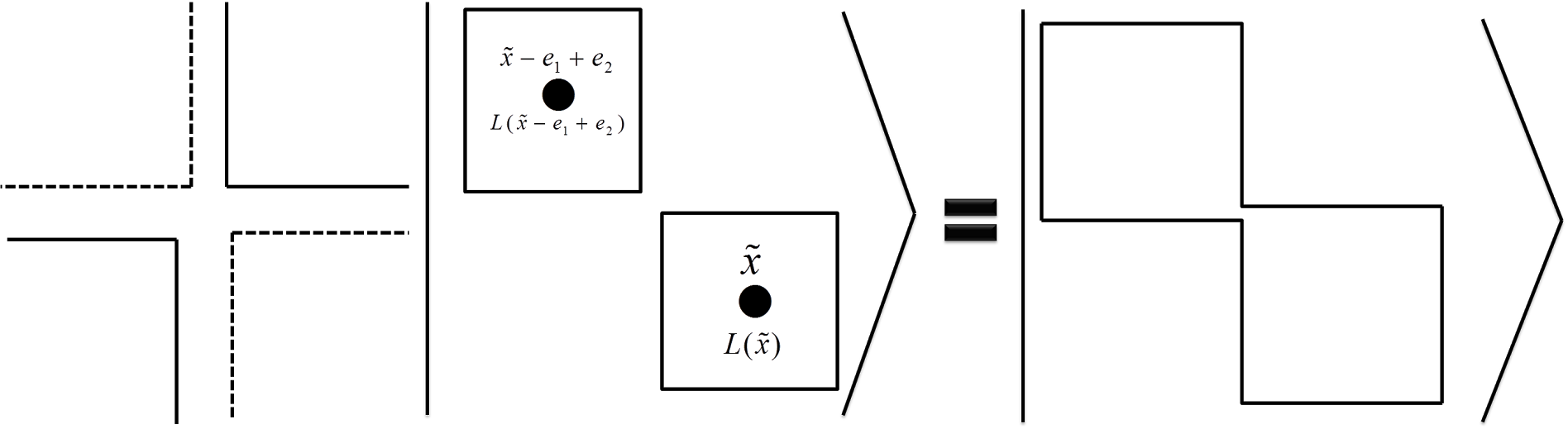}
\end{center}
\caption{$ \Pi^-_{D_2}\Pi^+_{D_1}(\tilde x -\frac{e_1}{2}+\frac{e_2}{2})|L(\tilde x)=1,L(\tilde x-e_1+e_2)=1\rangle= |L(\tilde x)=1,L(\tilde x+e_1)=1,{D_1}(\tilde x -\frac{e_1}{2}=1,{D_2}(\tilde x -\frac{e_1}{2}=-1\rangle$} 
\label{eg3}
\end{figure}
\end{itemize}
\item The quantum number $L$ counts the flux around a plaquette, hence it is natural to assign the variable $L$ at the centre of each plaquette, i.e at each dual site by defining $L(\tilde x)$, where, $\tilde x=x+ \frac{e_1}{2}+ \frac{e_2}{2}$. Similarly, as shown in figure \ref{lnq}, we can naturally assign the variable $N_2$ to the midpoint of each for the vertical links, i.e $N_2(\tilde x-\frac{e_1}{2})$ and the variable $N_1$ to the midpoint of each for the horizontal links, i.e $N_1(\tilde x-\frac{e_2}{2})$. The variables $D_{1(2)}$ are naturally assigned to each original lattice site, i.e $D_{1(2)}(\tilde x-\frac{e_1}{2}- \frac{e_2}{2})$. This particular set of quantum numbers defined at and around a dual lattice site, is sufficient to characterize any loops in the theory, or in other words, each and every loops of the theory can be uniquely specified by specifying a set of fusion quantum numbers locally throughout the lattice. 
\item We have already seen in the above example that, the basic loop variable $L$ can take any positive value and is independent of others. However, the other variables can be both positive and negative but are defined within a finite range. These new set of fusion variables are related to the linking quantum numbers in the following way: 
\bea
\label{mr1}
l_{12}(x)&=& L(\tilde x)-N_2(\tilde x-\frac{e_1}{2})-N_1(\tilde x-\frac{e_2}{2})+D_1(\tilde x-\frac{e_1}{2}-\frac{e_2}{2})\ge 0\\
\label{mr2}
l_{1\bar 1}(x)&=& N_2(\tilde x-\frac{e_1}{2})+ N_2(\tilde x-\frac{e_1}{2}-e_2)-D_1(\tilde x-\frac{e_1}{2}-\frac{e_2}{2})-D_2(\tilde x-\frac{e_1}{2}-\frac{e_2}{2})\ge 0\\
\label{mr3}
l_{1\bar 2}(x)&=& L(\tilde x-e_2)-N_2(\tilde x-\frac{e_1}{2}-e_2)-N_1(\tilde x-\frac{e_2}{2})+D_2(\tilde x-\frac{e_1}{2}-\frac{e_2}{2})\ge 0\\
\label{mr4}
l_{2\bar 1}(x)&=& L(\tilde x-e_1)-N_2(\tilde x-\frac{e_1}{2})-N_1(\tilde x-e_1-\frac{e_2}{2})+D_2(\tilde x-\frac{e_1}{2}-\frac{e_2}{2})\ge 0\\
\label{mr5}
l_{2\bar 2}(x)&=&N_1(\tilde x-\frac{e_2}{2})+ N_1(\tilde x-e_1-\frac{e_2}{2})-D_1(\tilde x-\frac{e_1}{2}-\frac{e_2}{2})-D_2(\tilde x-\frac{e_1}{2}-\frac{e_2}{2})\ge 0\\
\label{mr6}
l_{\bar 1\bar 2}(x)&=& L(\tilde x-e_1-e_2)-N_2(\tilde x-\frac{e_1}{2}-e_2)-N_1(\tilde x-e_1-\frac{e_2}{2})+D_1(\tilde x-\frac{e_1}{2}--\frac{e_2}{2})\ge 0
\eea
The above set of relations can be realized easily from figure \ref{lnq}. The fusion quantum numbers can take any positive or negative value over the lattice but the right hand sides of the set of equations (\ref{mr1}-\ref{mr6}), i.e the linking quantum numbers must always be positive semi-definite. This imposes a quite non-trivial boundary condition for the allowed range of fusion quantum numbers.
\item   For any arbitrary loop, the number of prepotentials on each link of the lattice are counted following (\ref{n1},\ref{n2},\ref{n3},\ref{n4}) as:
\bea
\label{nnew1}
n_1(x)&=& L(\tilde x)+L(\tilde x-e_2)-2N_1(\tilde x-\frac{e_2}{2})~=~ n_{\bar 1}(x+e_1)\\
\label{nnew2}
n_2(x) &=& L(\tilde x)+L(\tilde x-e_1)-2N_2(\tilde x-\frac{e_1}{2})~=~ n_{\bar 2}(x+e_2)
\eea
Note that, the U(1) constraints are automatically satisfied in (\ref{nnew1}) and (\ref{nnew2}). 
\item Note that, the description of local linking states in terms of five linking numbers provides a complete description of loop states corresponding to only the physical degrees of freedom of the theory subject to the Mandelstam constrain together with the two U(1) constraint. The equivalent description of loop states in terms of five fusion loop numbers are again complete. Here, the U(1) constraints are solved trivially by construction, hence after solving the Mandelstam constraint one is left with four degrees of freedom implying that there exists another constraint in these variables which needs to be imposed to get the exact and complete loop basis. We will discuss that extra constraint later in this section.
\end{itemize}

From these construction, we can label the loop states as $|L,N_1,N_2,D_1,D_2\rangle$, which  are eigenstates of the following operators with the corresponding eigenvalues:
\bea
\hat L(\tilde x)|L,N_1,N_2,D_1,D_2\rangle &=& L(\tilde x)|L,N_1,N_2,D_1,D_2\rangle\nonumber \\
\hat N_1(\tilde x- \frac{e_2}{2})|L,N_1,N_2,D_1,D_2\rangle &=& N_1(\tilde x- \frac{e_2}{2})|L,N_1,N_2,D_1,D_2\rangle\nonumber \\
\hat N_2(\tilde x- \frac{e_1}{2})|L,N_1,N_2,D_1,D_2\rangle &=& N_2(\tilde x- \frac{e_1}{2})|L,N_1,N_2,D_1,D_2\rangle\nonumber \\
\hat D_1(\tilde x- \frac{e_1}{2}- \frac{e_2}{2})|L,N_1,N_2,D_1,D_2\rangle &=& D_1(\tilde x- \frac{e_1}{2}- \frac{e_2}{2})|L,N_1,N_2,D_1,D_2\rangle\nonumber \\
\hat D_2(\tilde x- \frac{e_1}{2}- \frac{e_2}{2})|L,N_1,N_2,D_1,D_2\rangle &=& D_2(\tilde x- \frac{e_1}{2}- \frac{e_2}{2})|L,N_1,N_2,D_1,D_2\rangle\nonumber \\
\eea
and the shift operators $\Pi^{\pm}$ corresponding to each of the fusion variables are defined by,
\bea
\hat L(\tilde x)\Pi^{\pm}_L(\tilde x)|L,N_1,N_2,D_1,D_2\rangle &=& \Big(L(\tilde x){\pm}1\Big)|L,N_1,N_2,D_1,D_2\rangle\nonumber \\
\hat N_1(\tilde x- \frac{e_2}{2})\Pi^{\pm}_{N_1}(\tilde x- \frac{e_2}{2})|L,N_1,N_2,D_1,D_2\rangle &=& \Big(N_1(\tilde x- \frac{e_2}{2}){\pm}1\Big)|L,N_1,N_2,D_1,D_2\rangle\nonumber \\
\hat N_2(\tilde x- \frac{e_1}{2})\Pi^{\pm}_{N_2}(\tilde x- \frac{e_1}{2})|L,N_1,N_2,D_1,D_2\rangle &=& \Big(N_2(\tilde x- \frac{e_1}{2}){\pm}1\Big)|L,N_1,N_2,D_1,D_2\rangle\nonumber \\
\hat D_1(\tilde x- \frac{e_1}{2}- \frac{e_2}{2})\Pi^{\pm}_{D_1}(\tilde x- \frac{e_1}{2}- \frac{e_2}{2})|L,N_1,N_2,D_1,D_2\rangle &=&\Big( D_1(\tilde x- \frac{e_1}{2}- \frac{e_2}{2}){\pm}1\Big)|L,N_1,N_2,D_1,D_2\rangle\nonumber \\
\hat D_2(\tilde x- \frac{e_1}{2}- \frac{e_2}{2})\Pi^{\pm}_{D_2}(\tilde x- \frac{e_1}{2}- \frac{e_2}{2})|L,N_1,N_2,D_1,D_2\rangle &=& \Big(D_2(\tilde x- \frac{e_1}{2}- \frac{e_2}{2}){\pm}1\Big)|L,N_1,N_2,D_1,D_2\rangle\nonumber \\\label{pi+}
\eea

It is evident from (\ref{nnew1}) and (\ref{nnew2}) that, the number operator constraints (\ref{noc}) present in prepotential formulation are already solved by the fusion variables.
 However, the fusion variable are five in number in contrast to only three physical degrees of freedom. This implies that  there still exist two constraints to be imposed on the Hilbert space of states characterized by fusion variables to obtain the physical loop space. We will discuss about those constraints in the next section.

\begin{figure}[h]
\begin{center}
\includegraphics[width=0.3\textwidth,height=0.3\textwidth]
{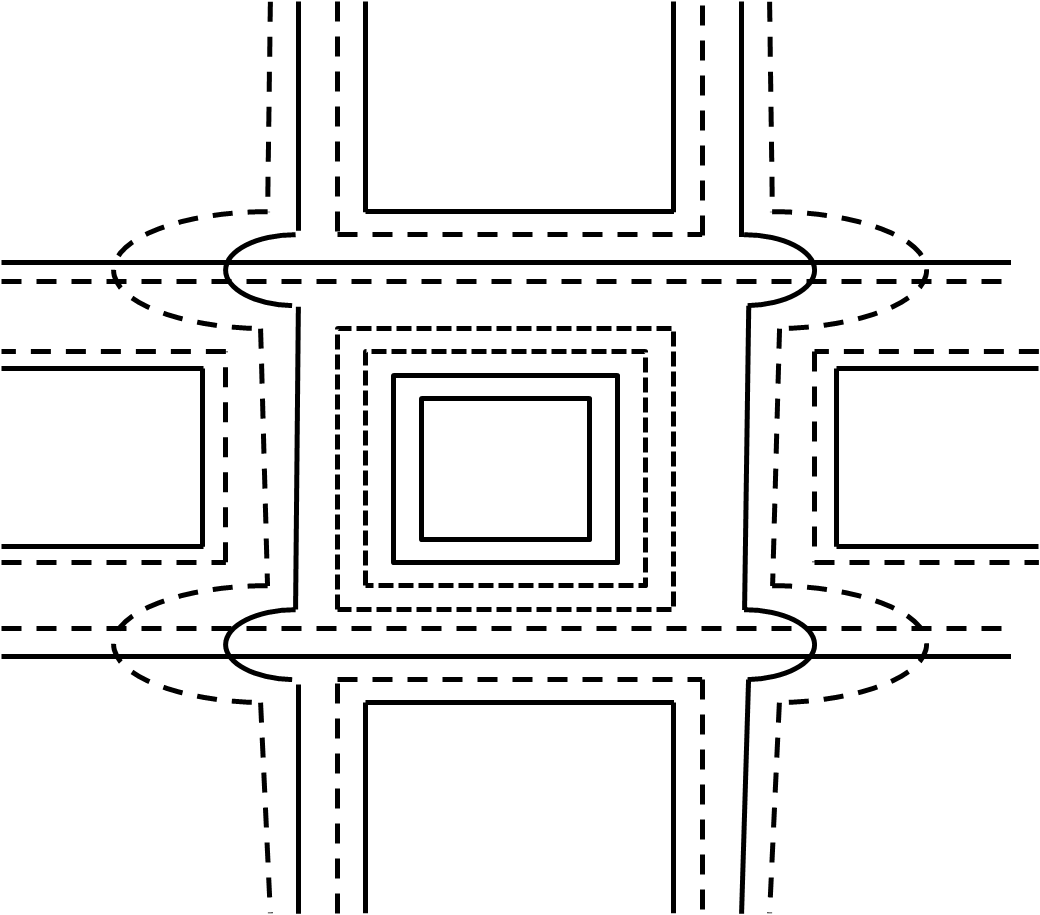}
\end{center}
\caption{Pictorial representation of the fusion constraint as given in (\ref{add_cons}).} 
\label{fc}
\end{figure}

The Mandelstam constraints are already solved when we consider our loop Hilbert space consisting of only non-intersecting loops by explicitly imposing:
\bea
l_{1\bar 1}(x)l_{2\bar 2}(x)&\equiv & \left( N_2(\tilde x-\frac{e_1}{2})+ N_2(\tilde x-\frac{e_1}{2}-e_2)-D_1(\tilde x-\frac{e_1}{2}-\frac{e_2}{2})-D_2(\tilde x-\frac{e_1}{2}-\frac{e_2}{2}) \right) \nonumber \\&&\left( N_1(\tilde x-\frac{e_2}{2})+ N_1(\tilde x-e_1-\frac{e_2}{2})-D_1(\tilde x-\frac{e_1}{2}-\frac{e_2}{2})-D_2(\tilde x-\frac{e_1}{2}-\frac{e_2}{2}) \right)=0~~~~\label{mcfn}
\eea
As stated earlier, apart from the constraint (\ref{mcfn}), there still exists another constraint in the fusion quantum number characterization of loop state in order to obtain three physical degrees of freedom. This additional constraint, which we name ``fusion constraint'' and is given by:
\bea
\label{add_cons}
&&\Pi_{D_2}^-(\tilde x-\frac{e_1}{2}-\frac{e_2}{2})\Pi_{D_2}^-(\tilde x+\frac{e_1}{2}+\frac{e_2}{2})\Pi_{D_1}^-(\tilde x-\frac{e_1}{2}+\frac{e_2}{2})\Pi_{D_1}^-(\tilde x+\frac{e_1}{2}-\frac{e_2}{2})\nonumber \\ &&
\Pi_{N_1}^+(\tilde x-\frac{e_2}{2})\Pi_{N_1}^+(\tilde x+\frac{e_2}{2})\Pi_{N_2}^+(\tilde x-\frac{e_1}{2})\Pi_{N_2}^+(\tilde x+\frac{e_1}{2})\left( \Pi^+_L(\tilde x) \right)^2=1
\eea
This fusion constraint is shown diagrammatically in figure \ref{fc}.
Note that, the fusion constraint and the Mandelstam constraint discussed earlier are independent of each other and hence commutes amongst themselves.

In the next section we write the Hamiltonian in terms of the Fusion variables.

\section{Hamiltonian Dynamics}
The dynamics of loop states under Kogut-Susskind Hamiltonian given in (\ref{ham}), can be realized in terms of Fusion variables as well as the corresponding shift operators we have defined earlier. 
In this section, we consider the Hamiltonian operator and its action on loop states characterized by fusion variables. The Hamiltonian for lattice gauge theory given in (\ref{ham}) consists of two parts. The electric part of the Hamiltonian which becomes dominant in the strong coupling limit of the theory, measures the flux along all the links of the lattice, whereas the magnetic part of the Hamiltonian, which is dominant in the weak coupling limit of the theory, is responsible for the dynamics of the loop states. 

The electric part of the Hamiltonian counts the total SU(2) flux on all the links of the lattice, which in terms of prepotential number operator is given by,
\bea
\label{ham1}
\hat H_{e}= g^2 \sum_{\mbox{links}} E^{2}_{links} = g^2 \sum_{x} \left[ \frac{n_1(x)}{2}\left( \frac{n_1(x)}{2}+1\right)+\frac{n_2(x)}{2}\left( \frac{n_2(x)}{2}+1\right)\right] \eea
where, $n_1(x)$ and $n_2(x)$ are the eigenvalues of the total number operator $\hat n$ counting the number of prepotentials (left or right) on the links along $1$ and $2$ directions originating at the site $x$. In terms of fusion operators, the total flux along the two links at each site are counted as given in (\ref{nnew1}) and (\ref{nnew2}). Using that, the electric part of the Hamiltonian  is given by:
\bea
\label{ham11}
\hat H_{e} &=& g^2 \sum_{\tilde x} \Bigg[ \left(\frac{L(\tilde x)+L(\tilde x-e_2)-2N_1(\tilde x-\frac{e_2}{2})}{2}\right)\left( \frac{L(\tilde x)+L(\tilde x-e_2)-2N_1(\tilde x-\frac{e_2}{2})}{2}+1\right)\nonumber \\&& +\left(\frac{L(\tilde x)+L(\tilde x-e_1)-2N_2(\tilde x-\frac{e_1}{2})}{2}\right)\left( \frac{L(\tilde x)+L(\tilde x-e_1)-2N_2(\tilde x-\frac{e_1}{2}) }{2}+1\right)\Bigg]
 \eea
Now we concentrate on the magnetic part given by,
\bea
\label{ham2}
H_{mag}=\frac{1}{g^2}\left(\mbox{Tr} U_{plaquette}+\mbox{Tr} U^\dagger_{plaquette}\right)
\eea
This is not as simple as the electric part even in terms of prepotentials or fusion variables.
 The magnetic Hamiltonian contains the gauge invariant loop operators. In previous sections we have already studied the actions of loop operators on loop states and have developed a diagrammatic technique to realize these actions which we will utilize now to find the action of the magnetic part of the Hamiltonian on any arbitrary loop state. Note that, we will consider the loop Hilbert space to contain only those states which solves the Mandelstam constraint, i.e satisfies (\ref{mcsol}). 

\begin{figure}[h]
\begin{center}
\includegraphics[width=0.8\textwidth,height=0.6\textwidth]
{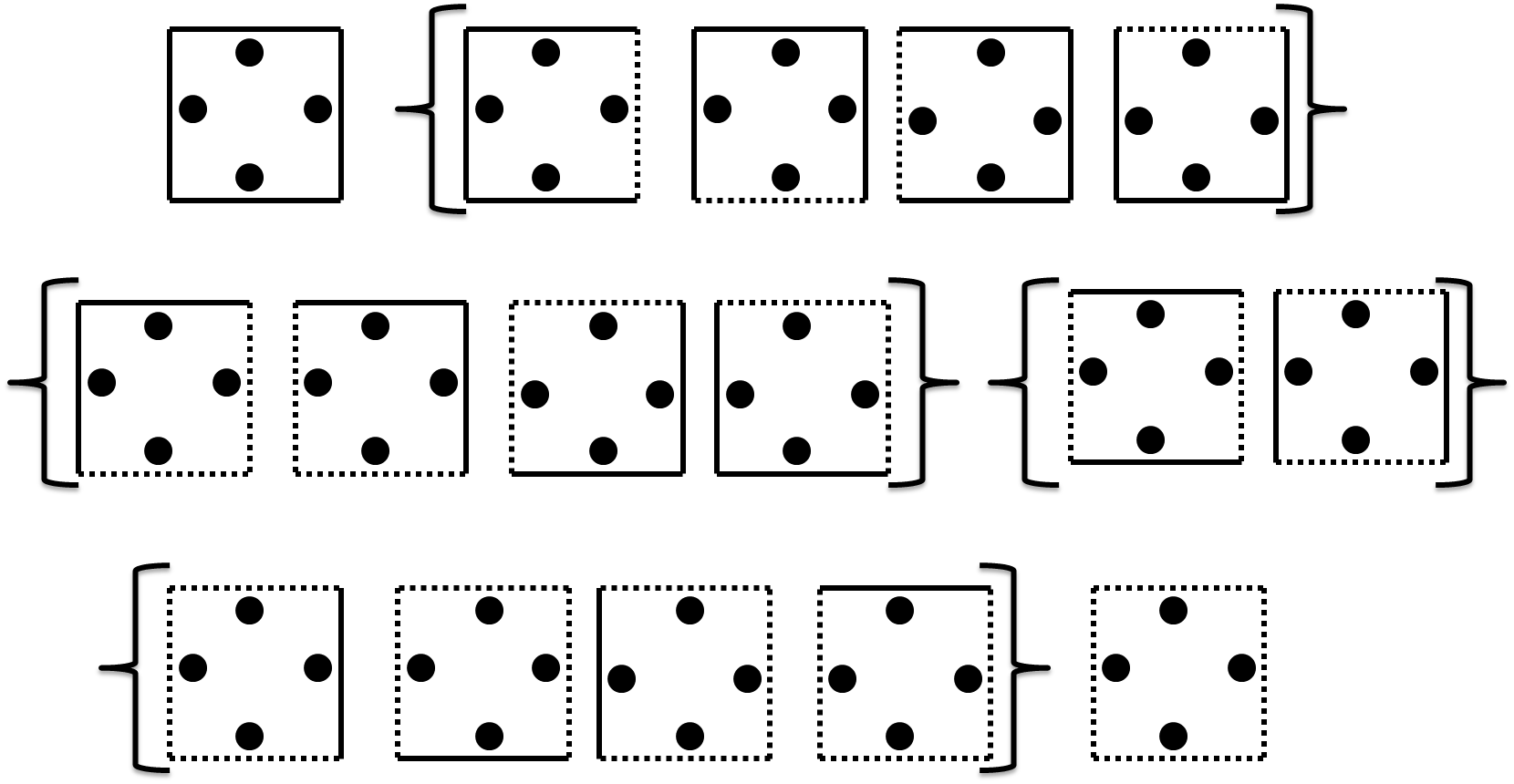}
\end{center}
\caption{The Hamiltonian operator in terms of prepotential becomes a sum of sixteen operators as shown above diagrammatically. The solid line along a link denotes the presence of prepotential creation operator on that link whereas, a dotted line denotes the annihilation operators on that. Clearly the whole set is rotationally symmetric and hermitian. These set of operators again can be subdivided in six classes of operators as shown in (a), (b), (c), (d), (e) and (f) denoting by (a)$\equiv H_{++++}$, (b)$\equiv H_{+++-}$, (c)$\equiv H_{++--}$, (d)$\equiv H_{+-+-}$, (e)$\equiv H_{+---}$, (f)$\equiv H_{----}$.} 
\label{h16}
\end{figure}

In terms of prepotentials each link operator breaks into $2$ parts as given in (\ref{su2U}). One of these two parts contains only the creation operator and the other only annihilation, making $U\equiv U_++U_-$. Hence,  the prepotential formulation enables us to write down the gauge invariant plaquette operator, which is trace of the products of four link operators around a plaquette, as a sum of $2^4=16$ operators as shown in figure \ref{h16}.
The constituent operators fall among different classes. We analyze each class separately and calculate the dynamics of physical loop states in each case. 
 Each of these plaquette operators are basically product of four different local gauge invariant operators at the four vertices. We have already studied these individual loop operators and have found their actions in (\ref{ac++},\ref{ac+-},\ref{ac--}). Now we exploit those calculations to compute the combinations of loop states produced by the action of the Hamiltonian.

Mandelstam constraint, (\ref{mcsol}) implies, that in the action of the loop operator $\mathcal O^{i_-j_-}$, as shown in figure \ref{ac--fig}, the last diagram of the right hand side would vanish. Hence within the loop space we consider, we will have the reduced action for the loop operators. Let us now consider the actions of each plaquette operators individually:
\begin{enumerate}
\item 
\label{type1}
The operator in (a) of figure \ref{h16} is $H_{++++}$. The local loop operators at each vertex are $\mathcal O^{i_+j_+}$ which acts according figure \ref{ac++fig}, yielding only one loop state as shown in figure \ref{h0-}. 
\begin{figure}[h]
\begin{center}
\includegraphics[width=0.2\textwidth,height=0.1\textwidth]
{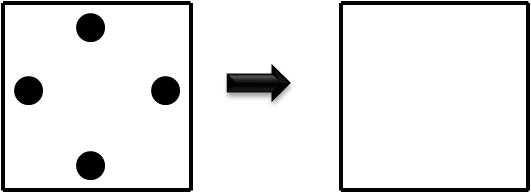}
\end{center}
\caption{Explicit action of type (a) or $H_{++++}$} 
\label{h0-}
\end{figure}
\item\label{type2}
The operator of type (b) are $H_{+++-}$, where at two adjacent vertices, the loop operators are $\mathcal O^{i_+j_+}$ giving rise to only one state, and at the opposite two they are of the type $\mathcal O^{i_+j_-}$ giving rise to $2\times 2$ states following figure \ref{ac+-fig}. Hence each plaquette operators of type (b) deforms the loop states on which it acts in $1\times 1\times 2\times 2=4$ possible way as shown in figure  \ref{h1-}. There are 4 such operators in type (b), which gives a total of 16 loop states. 
\begin{figure}[h]
\begin{center}
\includegraphics[width=0.4\textwidth,height=0.2\textwidth]
{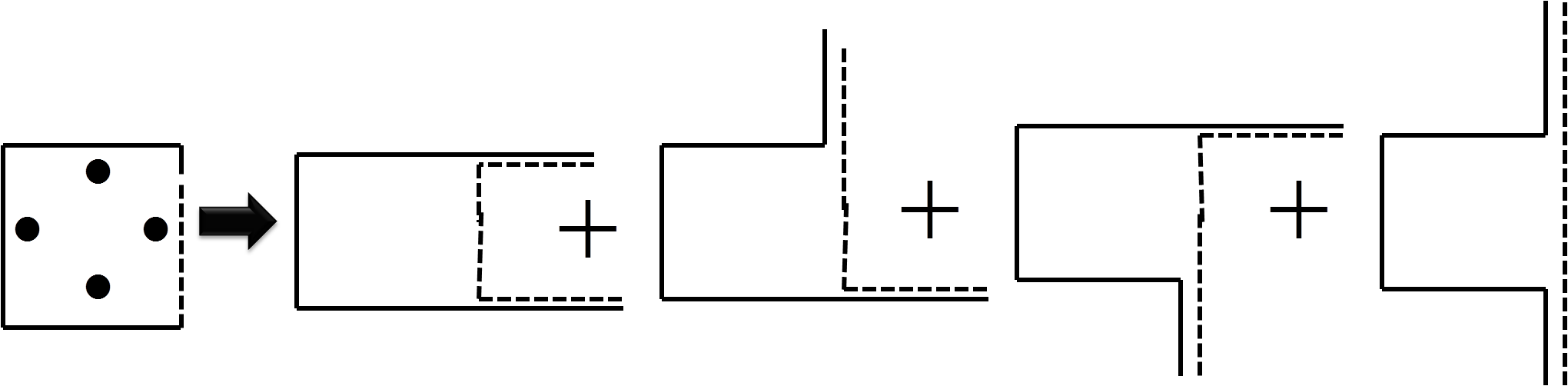}
\end{center}
\caption{Explicit action of  type (b) or  $H_{+++-}$} 
\label{h1-}
\end{figure}
\item\label{type3}
The (c) type operators, $H_{++--}$, where at one vertex the loop operator is $\mathcal O^{i_+j_+}$   and at the diagonally opposite vertex it is $\mathcal O^{i_-j_-}$. The first one gives only one loop state whereas the second one generates two following figure \ref{ac--fig} (NOT $3$ for loops which satisfy Mandelstam constraint). The other two vertices  are of the type $\mathcal O^{i_+j_-}$ giving rise to $2\times 2$ states following figure \ref{ac+-fig}. Hence each plaquette operators of type (c) deforms the loop states on which it acts in $1\times 2\times 2\times 2=8$ possible way as shown in figure  \ref{h2-1} and there are 4 such plaquette operators present. 
\begin{figure}[h]
\begin{center}
\includegraphics[width=0.6\textwidth,height=0.3\textwidth]
{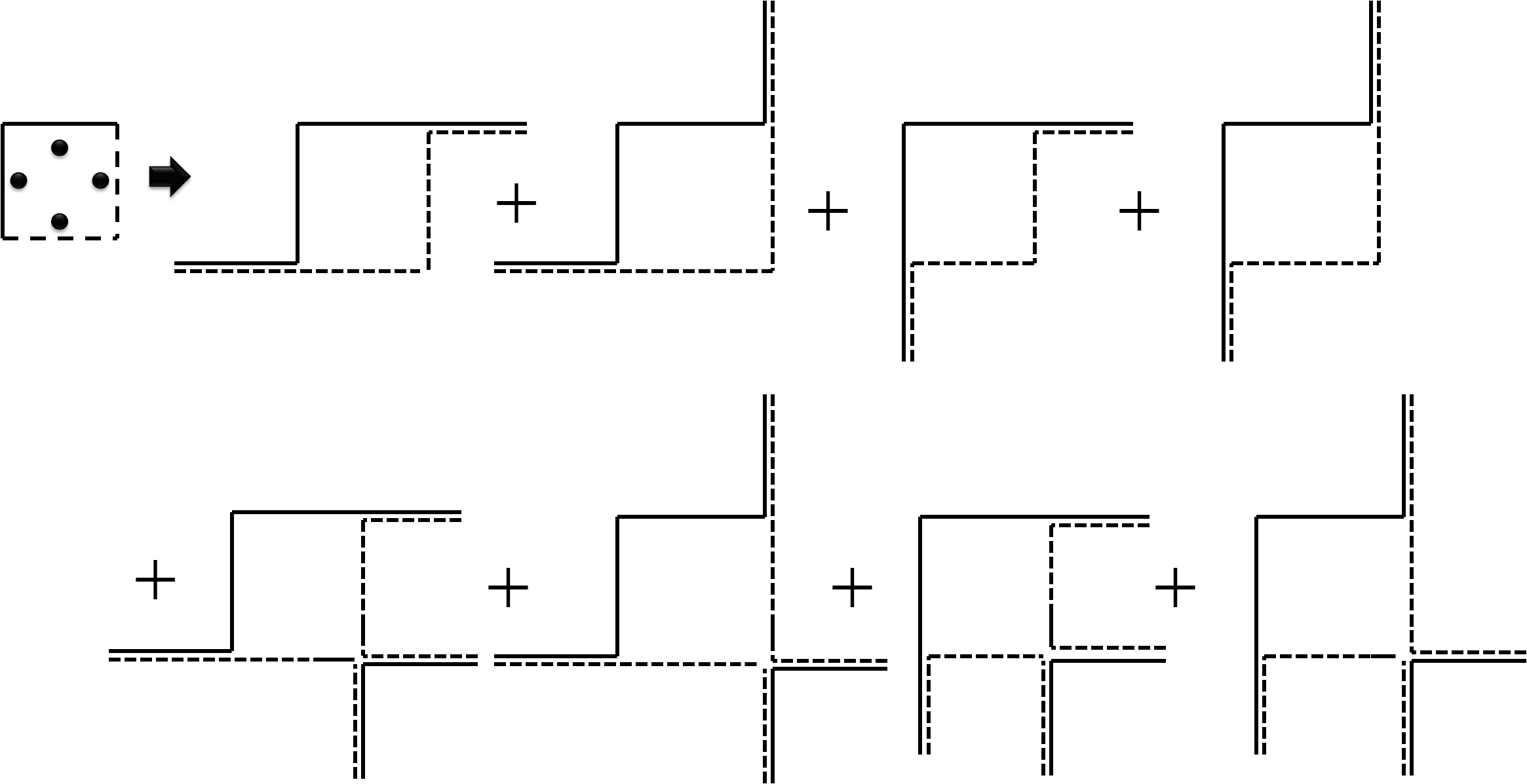}
\end{center}
\caption{Explicit action of  type (c) or $H_{++--}$} 
\label{h2-1}
\end{figure}
\item\label{type4}
The action of operators of type (d), i.e $H_{+-+-}$, are obtained by using figure \ref{ac+-fig} for the operators of type $\mathcal O^{i_+j_-}$ at all four vertices, yielding total of $2^4=16$ terms for each of the two such operators present in the class. The explicit states are given in figure \ref{h2-2}.
\begin{figure}[h]
\begin{center}
\includegraphics[width=0.5\textwidth,height=0.3\textwidth]
{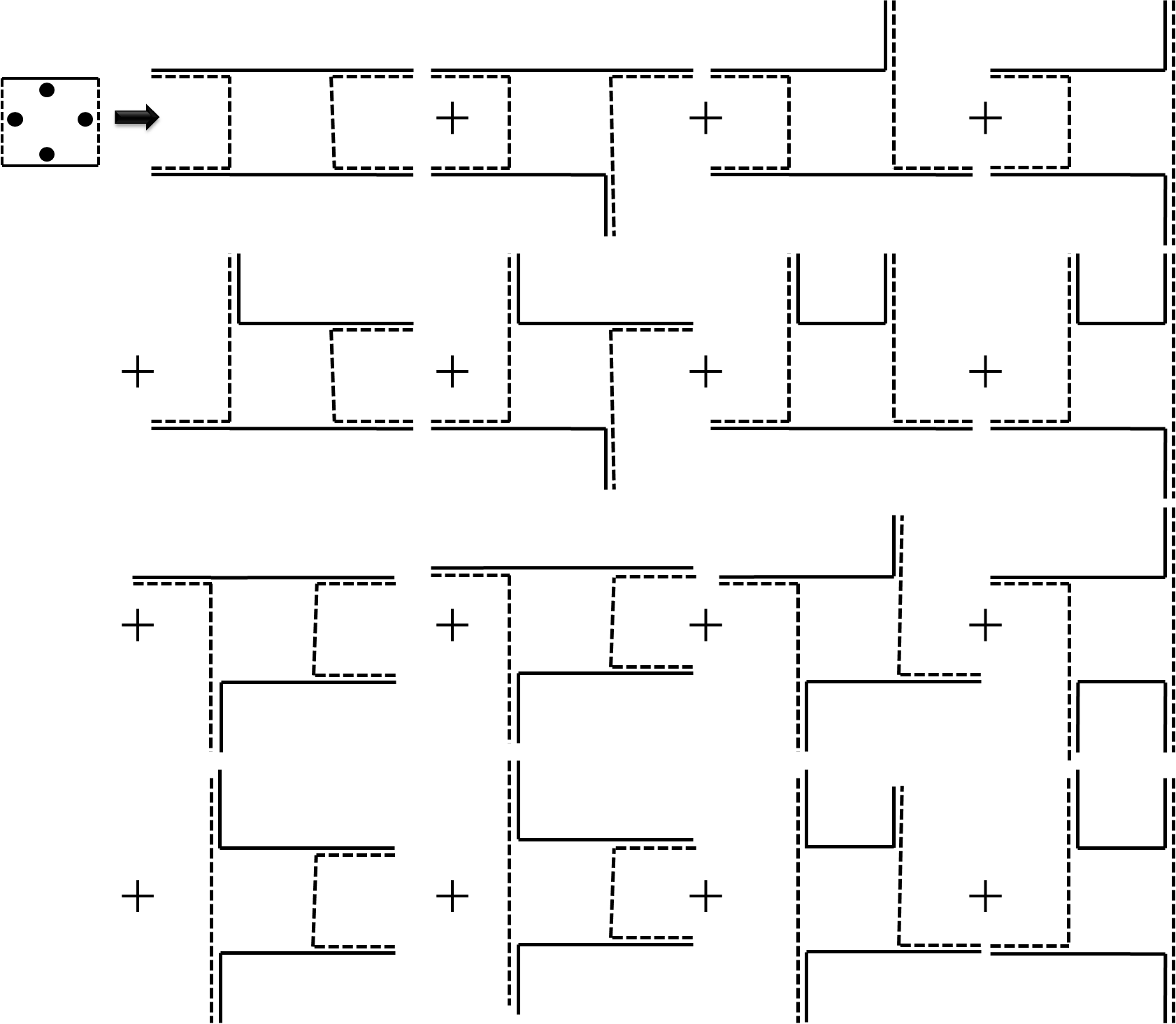}
\end{center}
\caption{Explicit action of type (d) or  $H_{+-+-}$} 
\label{h2-2}
\end{figure}
\item\label{type5}
The operator of type (e) are $H_{+---}$, where at two adjacent vertices, the loop operators are $\mathcal O^{i_-j_-}$ giving rise to $2$ states each following figure \ref{ac--fig}, and  the opposite two they are of the type $\mathcal O^{i_+j_-}$ again giving rise to $2$ states each following figure \ref{ac+-fig}. Hence each plaquette operators of type (e) deforms the loop states on which it acts in $2^4=16$ possible way as shown in figure  \ref{h3-}. There are $4$ such operators in type (e), which gives a total of $64$ loop states. 
\begin{figure}[h]
\begin{center}
\includegraphics[width=0.6\textwidth,height=0.8\textwidth]
{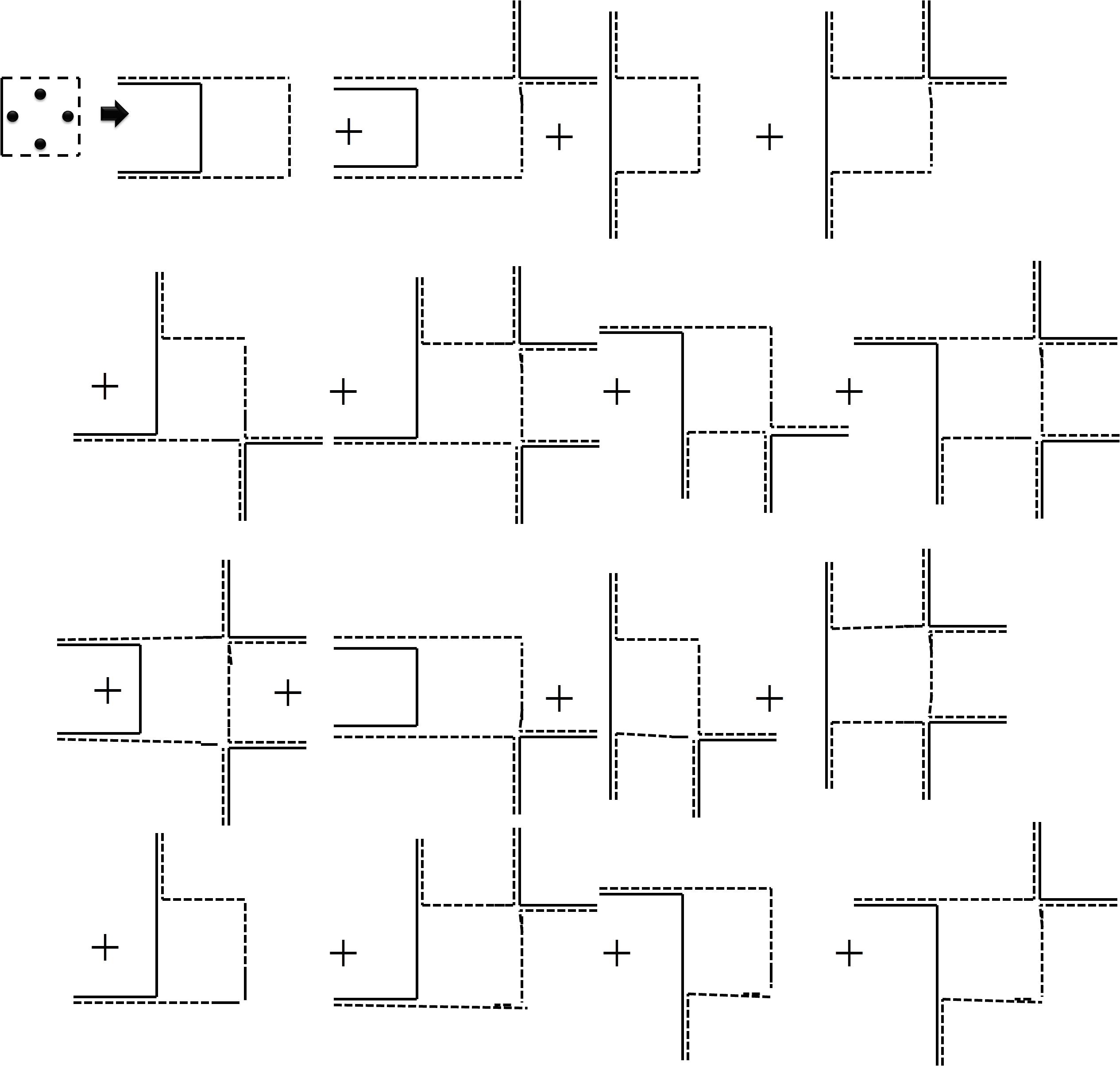}
\end{center}
\caption{Explicit action of  type (e) or $H_{+---}$} 
\label{h3-}
\end{figure}
\item\label{type6}
Finally, for type (f), i.e $H_{----}$ at all the four vertices the loop operators are $\mathcal O^{i_-j_-}$ giving rise to $2$ states each following figure \ref{ac--fig}, yielding $2^4=16$ loop states as shown in  figure \ref{h4-}.
\begin{figure}[h]
\begin{center}
\includegraphics[width=0.6\textwidth,height=0.3\textwidth]
{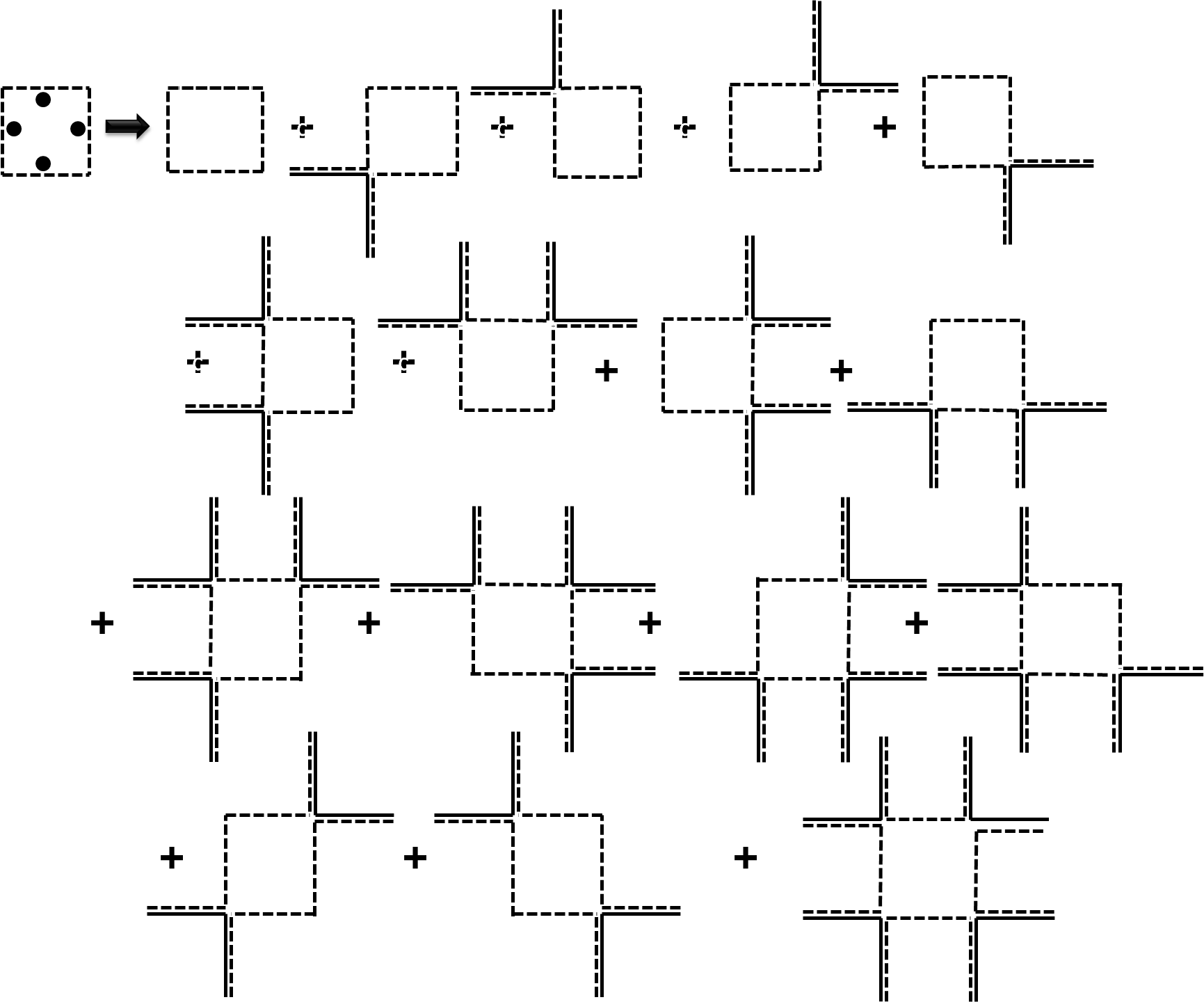}
\end{center}
\caption{Explicit action of  type (f) or $H_{----}$} 
\label{h4-}
\end{figure}
\end{enumerate}
The loop states produced by the action of the magnetic part of the Hamiltonian as discussed so far, can also be realized to be created by the actions of the shift operators $\Pi^\pm$ corresponding to the fusion variables as given in (\ref{pi+}) together with a particular coefficient associated and fixed by each diagram.
The action of the Hamiltonian on loop states has been described in figures \ref{h0-}, \ref{h1-}, \ref{h2-1}, \ref{h2-2}, \ref{h3-}, \ref{h4-}. These diagrams denotes that for each loop state created, the fusion new state can be realized by a new set of fusion quantum numbers. Or in other words, the Hamiltonian can be represented by  shift operators in fusion variables together with the a certain coefficient which describes the new state created. Using the diagrammatic rules provided in  figure \ref{rule} and the equations thereafter, we can calculate that coefficient. 
Now, from each diagram in figures \ref{h0-}, \ref{h1-}, \ref{h2-1}, \ref{h2-2}, \ref{h3-}, \ref{h4-}, one can read the constant coefficient in front of it, and the change in fusion quantum numbers for each term. The action of the magnetic Hamiltonian on loop states $|L,N_1,N_2,D_1,D_2\rangle$ is thus obtained as:
\bea
\mbox{Type (a):}&&\nonumber \\
H_1 &=& C^{1_+2_+}_a C^{2_+\bar 1_+}_b C^{\bar 1_+\bar 2_+}_c C^{1_+\bar 2_+}_d \Pi_L^+(\tilde x)  
\\
\mbox{Type (b):}&&\nonumber \\
H_2 &=& C^{1_+2_+}_aC^{1_+\bar 2_+}_d \left( (C^{\bar 1_+\bar 2_-}_c)_1 +(C^{\bar 1_+\bar 2_-}_c)_2 \Pi^+_{D_2}(\tilde x+\frac{e_1}{2}+\frac{e_2}{2}) \right)\nonumber \\ &&\left( (C^{2_-\bar 1_+}_b)_{1}+(C^{2_-\bar 1_+}_b)_{\bar 2} \Pi^+_{D_1}(\tilde x+\frac{e_1}{2}-\frac{e_2}{2}) \right)\Pi^+_{N_2}(\tilde x+\frac{e_1}{2}) \Pi_L^+(\tilde x) \\
H_3 &=& C^{\bar 1_+\bar 2_+}_c C^{1_+\bar 2_+}_d  \left( (C^{1_-2_+}_a)_{\bar 2}+ (C^{1_-2_+}_a)_{\bar 1}\Pi^+_{D_2}(\tilde x-\frac{e_1}{2}-\frac{e_2}{2}) \right)\nonumber \\ && \left((C^{2_+\bar 1_-}_b)_{\bar 2}+ (C^{2_+\bar 1_-}_b)_{1}\Pi^+_{D_1}(\tilde x+\frac{e_1}{2}-\frac{e_2}{2}) \right)\Pi^+_{N_1}(\tilde x-\frac{e_2}{2}) \Pi_L^+(\tilde x) \\
H_4 &=&  C^{2_+\bar 1_+}_b C^{\bar 1_+\bar 2_+}_c \left( ( C^{1_+2_-}_a )_{\bar 1}+ ( C^{1_+2_-}_a )_{\bar 2}\Pi^+_{D_2}(\tilde x-\frac{e_1}{2}-\frac{e_2}{2}) \right)\nonumber \\ &&\left((C^{1_+\bar 2_-}_d  )_{\bar 1}+ ( C^{1_+\bar 2_-}_d )_2\Pi^+_{D_1}(\tilde x-\frac{e_1}{2}+\frac{e_2}{2}) \right)\Pi^+_{N_2}(\tilde x-\frac{e_1}{2}) \Pi_L^+(\tilde x) \\
H_5 &=&   C^{1_+2_+}_a C^{2_+\bar 1_+}_b \left( (C^{\bar 1_-\bar 2_+}_c  )_2+( C^{\bar 1_-\bar 2_+}_c )_1 \Pi^+_{D_2}(\tilde x+\frac{e_1}{2}+\frac{e_2}{2}) \right)\nonumber \\ &&
\left((  C^{1_-\bar 2_+}_d )_2+(  C^{1_-\bar 2_+}_d )_{\bar 1} \Pi^+_{D_1}(\tilde x-\frac{e_1}{2}+\frac{e_2}{2}) \right)\Pi^+_{N_1}(\tilde x+\frac{e_2}{2}) \Pi_L^+(\tilde x) \eea

\bea
\mbox{Type (c):}&&\nonumber \\
H_6 &=&  \left( C^{1_+\bar 2_+}_d\Pi^-_{D_1}(\tilde x-\frac{e_1}{2}+\frac{e_2}{2}) \right)\left( (C^{\bar 1_+\bar 2_-}_c)_1 +(C^{\bar 1_+\bar 2_-}_c)_2 \Pi^-_{D_2}(\tilde x+\frac{e_1}{2}+\frac{e_2}{2}) \right)  \nonumber \\ &&
\left( (C^{1_-2_+}_a)_{\bar 2}+ (C^{1_-2_+}_a)_{\bar 1}\Pi^-_{D_2}(\tilde x-\frac{e_1}{2}-\frac{e_2}{2}) \right) \\ &&
 \left(C^{2_-\bar 1_-}_b+ C^{(2_-)_1(\bar 1_-)_{\bar 2}}_b\Pi^+_{D_2}(\tilde x+\frac{e_1}{2}-\frac{e_2}{2}) \Pi^-_{D_1}(\tilde x+\frac{e_1}{2}-\frac{e_2}{2})\right) \Pi^-_{N_2}(\tilde x-\frac{e_1}{2}) \Pi^-_{N_1}(\tilde x+\frac{e_2}{2}) \Pi_L^-(\tilde x)\nonumber \\ 
H_7 &=& \left( C^{\bar 1_+\bar 2_+}_c\Pi^-_{D_2}(\tilde x+\frac{e_1}{2}+\frac{e_2}{2}) \right) \left((C^{2_+\bar 1_-}_b)_{\bar 2}+ (C^{2_+\bar 1_-}_b)_{1}\Pi^-_{D_1}(\tilde x+\frac{e_1}{2}-\frac{e_2}{2}) \right)\nonumber\\ &&
\left((C^{1_+\bar 2_-}_d  )_{\bar 1}+ ( C^{1_+\bar 2_-}_d )_2\Pi^-_{D_1}(\tilde x-\frac{e_1}{2}+\frac{e_2}{2}) \right) \\ && 
\left(C^{1_-2_-}_a+C^{(1_-)_{\bar 2}(2_-)_{\bar 1}}_a \Pi^+_{D_1}(\tilde x-\frac{e_1}{2}-\frac{e_2}{2}) \Pi^-_{D_2}(\tilde x-\frac{e_1}{2}-\frac{e_2}{2})\right) \Pi^-_{N_2}(\tilde x+\frac{e_1}{2}) \Pi^-_{N_1}(\tilde x+\frac{e_2}{2}) \Pi_L^-(\tilde x)\nonumber \\
H_8 &=& \left( C^{2_+\bar 1_+}_b\Pi^-_{D_1}(\tilde x+\frac{e_1}{2}-\frac{e_2}{2}) \right) \left( (C^{\bar 1_+\bar 2_-}_c)_1 +(C^{\bar 1_+\bar 2_-}_c)_2 \Pi^-_{D_2}(\tilde x+\frac{e_1}{2}+\frac{e_2}{2}) \right) \nonumber \\ && \left( (C^{1_-2_+}_a)_{\bar 2}+ (C^{1_-2_+}_a)_{\bar 1}\Pi^-_{D_2}(\tilde x-\frac{e_1}{2}-\frac{e_2}{2}) \right)\\&&
 \left(C^{1_-\bar 2_-}_d + C^{(1_-)_2(\bar 2_-)_{\bar 1}}_d  \Pi^+_{D_2}(\tilde x-\frac{e_1}{2}+\frac{e_2}{2}) \Pi^-_{D_1}(\tilde x-\frac{e_1}{2}+\frac{e_2}{2})\right) \Pi^-_{N_2}(\tilde x+\frac{e_1}{2}) \Pi^-_{N_1}(\tilde x-\frac{e_2}{2}) \Pi_L^-(\tilde x)\nonumber \\
H_9 &=&  \left( C^{1_+2_+}_a \Pi^-_{D_2}(\tilde x-\frac{e_1}{2}-\frac{e_2}{2}) \right)\left((C^{1_+\bar 2_-}_d  )_{\bar 1}+ ( C^{1_+\bar 2_-}_d )_2\Pi^-_{D_1}(\tilde x-\frac{e_1}{2}+\frac{e_2}{2}) \right) \nonumber \\ && 
\left((C^{2_+\bar 1_-}_b)_{\bar 2}+ (C^{2_+\bar 1_-}_b)_{1}\Pi^-_{D_1}(\tilde x+\frac{e_1}{2}-\frac{e_2}{2}) \right)  \\ && 
\left(C^{\bar 1_-\bar 2_-}_c+C^{(\bar 1_-)_2(\bar 2_-)_{1}}_c \Pi^+_{D_1}(\tilde x+\frac{e_1}{2}+\frac{e_2}{2}) \Pi^-_{D_2}(\tilde x+\frac{e_1}{2}+\frac{e_2}{2})\right) \Pi^-_{N_2}(\tilde x-\frac{e_1}{2}) \Pi^-_{N_1}(\tilde x-\frac{e_2}{2}) \Pi_L^-(\tilde x)\nonumber
\eea

\bea
\mbox{Type (d):}&&\nonumber \\
H_{10} &=& \left( (C^{\bar 1_+\bar 2_-}_c)_1 +(C^{\bar 1_+\bar 2_-}_c)_2 \Pi^+_{D_2}(\tilde x+\frac{e_1}{2}+\frac{e_2}{2}) \right)\nonumber \\ &&\left( (C^{2_-\bar 1_+}_b)_{1}+(C^{2_-\bar 1_+}_b)_{\bar 2} \Pi^+_{D_1}(\tilde x+\frac{e_1}{2}-\frac{e_2}{2}) \right)\nonumber \\ && 
\left( ( C^{1_+2_-}_a )_{\bar 1}+ ( C^{1_+2_-}_a )_{\bar 2}\Pi^+_{D_2}(\tilde x-\frac{e_1}{2}-\frac{e_2}{2}) \right)\nonumber \\ &&\left((C^{1_+\bar 2_-}_d  )_{\bar 1}+ ( C^{1_+\bar 2_-}_d )_2\Pi^+_{D_1}(\tilde x-\frac{e_1}{2}+\frac{e_2}{2}) \right) \Pi^+_{N_2}(\tilde x+\frac{e_1}{2})\Pi^+_{N_2}(\tilde x-\frac{e_1}{2}) \Pi_L^+(\tilde x) \\
H_{11} &=& \left( (C^{1_-2_+}_a)_{\bar 2}+ (C^{1_-2_+}_a)_{\bar 1}\Pi^+_{D_2}(\tilde x-\frac{e_1}{2}-\frac{e_2}{2}) \right)\nonumber \\ && \left((C^{2_+\bar 1_-}_b)_{\bar 2}+ (C^{2_+\bar 1_-}_b)_{1}\Pi^+_{D_1}(\tilde x+\frac{e_1}{2}-\frac{e_2}{2}) \right) \nonumber \\ &&  \left( (C^{\bar 1_-\bar 2_+}_c  )_2+( C^{\bar 1_-\bar 2_+}_c )_1 \Pi^+_{D_2}(\tilde x+\frac{e_1}{2}+\frac{e_2}{2}) \right)\nonumber \\ &&
\left((  C^{1_-\bar 2_+}_d )_2+(  C^{1_-\bar 2_+}_d )_{\bar 1} \Pi^+_{D_1}(\tilde x-\frac{e_1}{2}+\frac{e_2}{2}) \right)  \Pi^+_{N_1}(\tilde x+\frac{e_2}{2})\Pi^+_{N_1}(\tilde x-\frac{e_2}{2}) \Pi_L^+(\tilde x) 
\eea\bea
\mbox{Type (e):}&&\nonumber \\
H_{12} &=& \left( (C^{\bar 1_+\bar 2_-}_c)_1 +(C^{\bar 1_+\bar 2_-}_c)_2 \Pi^-_{D_2}(\tilde x+\frac{e_1}{2}+\frac{e_2}{2}) \right)  \nonumber \\ &&
\left((C^{2_+\bar 1_-}_b)_{\bar 2}+ (C^{2_+\bar 1_-}_b)_{1}\Pi^-_{D_1}(\tilde x+\frac{e_1}{2}-\frac{e_2}{2}) \right)\nonumber\\ &&
\left(C^{1_-2_-}_a+C^{(1_-)_{\bar 2}(2_-)_{\bar 1}}_a \Pi^+_{D_1}(\tilde x-\frac{e_1}{2}-\frac{e_2}{2}) \Pi^-_{D_2}(\tilde x-\frac{e_1}{2}-\frac{e_2}{2})\right) \\ &&
\left(C^{1_-\bar 2_-}_d + C^{(1_-)_2(\bar 2_-)_{\bar 1}}_d  \Pi^+_{D_2}(\tilde x-\frac{e_1}{2}+\frac{e_2}{2}) \Pi^-_{D_1}(\tilde x-\frac{e_1}{2}+\frac{e_2}{2})\right)
\Pi^-_{N_2}(\tilde x+\frac{e_1}{2})  \Pi_L^-(\tilde x) \\
H_{13} &=&  \left( (C^{1_-2_+}_a)_{\bar 2}+ (C^{1_-2_+}_a)_{\bar 1}\Pi^-_{D_2}(\tilde x-\frac{e_1}{2}-\frac{e_2}{2}) \right)\nonumber \\&&
\left((C^{2_+\bar 1_-}_b)_{\bar 2}+ (C^{2_+\bar 1_-}_b)_{1}\Pi^-_{D_1}(\tilde x+\frac{e_1}{2}-\frac{e_2}{2}) \right)\nonumber \\ &&
\left(C^{\bar 1_-\bar 2_-}_c+C^{(\bar 1_-)_2(\bar 2_-)_{1}}_c \Pi^+_{D_1}(\tilde x+\frac{e_1}{2}+\frac{e_2}{2}) \Pi^-_{D_2}(\tilde x+\frac{e_1}{2}+\frac{e_2}{2})\right)    \nonumber \\ && 
 \left(C^{1_-\bar 2_-}_d + C^{(1_-)_2(\bar 2_-)_{\bar 1}}_d  \Pi^+_{D_2}(\tilde x-\frac{e_1}{2}+\frac{e_2}{2}) \Pi^-_{D_1}(\tilde x-\frac{e_1}{2}+\frac{e_2}{2})\right)     \Pi^-_{N_1}(\tilde x-\frac{e_2}{2}) \Pi_L^-(\tilde x)\\
H_{14} &=&   \left( (C^{1_-2_+}_a)_{\bar 2}+ (C^{1_-2_+}_a)_{\bar 1}\Pi^-_{D_2}(\tilde x-\frac{e_1}{2}-\frac{e_2}{2}) \right)\nonumber  \\ &&         
          \left((C^{1_+\bar 2_-}_d  )_{\bar 1}+ ( C^{1_+\bar 2_-}_d )_2\Pi^-_{D_1}(\tilde x-\frac{e_1}{2}+\frac{e_2}{2}) \right) \nonumber \\ && 
 \left(C^{2_-\bar 1_-}_b+ C^{(2_-)_1(\bar 1_-)_{\bar 2}}_b\Pi^+_{D_2}(\tilde x+\frac{e_1}{2}-\frac{e_2}{2}) \Pi^-_{D_1}(\tilde x+\frac{e_1}{2}-\frac{e_2}{2})\right)  \nonumber \\ && 
     \left(C^{\bar 1_-\bar 2_-}_c+C^{(\bar 1_-)_2(\bar 2_-)_{1}}_c \Pi^+_{D_1}(\tilde x+\frac{e_1}{2}+\frac{e_2}{2}) \Pi^-_{D_2}(\tilde x-\frac{e_1}{2}+\frac{e_2}{2})\right)            \Pi^-_{N_2}(\tilde x-\frac{e_1}{2}) \Pi_L^-(\tilde x)\\
H_{15} &=&  \left( (C^{\bar 1_+\bar 2_-}_c)_1 +(C^{\bar 1_+\bar 2_-}_c)_2 \Pi^-_{D_2}(\tilde x+\frac{e_1}{2}+\frac{e_2}{2}) \right)  \nonumber \\ &&  
   \left((C^{1_+\bar 2_-}_d  )_{\bar 1}+ ( C^{1_+\bar 2_-}_d )_2\Pi^-_{D_1}(\tilde x-\frac{e_1}{2}+\frac{e_2}{2}) \right)         \nonumber \\ &&
      \left(C^{1_-2_-}_a+C^{(1_-)_{\bar 2}(2_-)_{\bar 1}}_a \Pi^+_{D_1}(\tilde x-\frac{e_1}{2}-\frac{e_2}{2}) \Pi^-_{D_2}(\tilde x-\frac{e_1}{2}-\frac{e_2}{2})\right)   \nonumber \\ && 
         \left(C^{2_-\bar 1_-}_b+ C^{(2_-)_1(\bar 1_-)_{\bar 2}}_b\Pi^+_{D_2}(\tilde x+\frac{e_1}{2}-\frac{e_2}{2}) \Pi^-_{D_1}(\tilde x+\frac{e_1}{2}-\frac{e_2}{2})\right)        \Pi^-_{N_1}(\tilde x+\frac{e_2}{2}) \Pi_L^-(\tilde x)\\
\mbox{Type (f):}&&\nonumber \\
H_{16} &=&  \left(C^{1_-2_-}_a+C^{(1_-)_{\bar 2}(2_-)_{\bar 1}}_a \Pi^+_{D_1}(\tilde x-\frac{e_1}{2}-\frac{e_2}{2}) \Pi^-_{D_2}(\tilde x-\frac{e_1}{2}-\frac{e_2}{2})\right)   \nonumber \\ && 
         \left(C^{2_-\bar 1_-}_b+ C^{(2_-)_1(\bar 1_-)_{\bar 2}}_b\Pi^+_{D_2}(\tilde x+\frac{e_1}{2}-\frac{e_2}{2}) \Pi^-_{D_1}(\tilde x+\frac{e_1}{2}-\frac{e_2}{2})\right)  \nonumber \\ &&  \left(C^{\bar 1_-\bar 2_-}_c+C^{(\bar 1_-)_2(\bar 2_-)_{1}}_c \Pi^+_{D_1}(\tilde x+\frac{e_1}{2}+\frac{e_2}{2}) \Pi^-_{D_2}(\tilde x+\frac{e_1}{2}+\frac{e_2}{2})\right)    \nonumber \\ && 
 \left(C^{1_-\bar 2_-}_d + C^{(1_-)_2(\bar 2_-)_{\bar 1}}_d  \Pi^+_{D_2}(\tilde x-\frac{e_1}{2}+\frac{e_2}{2}) \Pi^-_{D_1}(\tilde x-\frac{e_1}{2}+\frac{e_2}{2})\right)   \Pi_L^-(\tilde x)
\eea
 In all the sixteen terms of the Hamiltonian, the coefficient $C$'s with suffix $a,b,c,d$ denotes them to be defined at points $(\tilde x-\frac{e_1}{2}-\frac{e_2}{2}),(\tilde x+\frac{e_1}{2}-\frac{e_2}{2}),(\tilde x+\frac{e_1}{2}+\frac{e_2}{2})$ and $(\tilde x-\frac{e_1}{2}+\frac{e_2}{2})$ respectively.

The matrix elements of this magnetic Hamiltonian within the loop states can be calculated following appendix B. In Appendix B we compute the norm of loop states by noticing that this is itself product of four norms defined at the four corner sites of a plaquette. In appendix C,  we briefly illustrate how the strong coupling series in this new formalism, using the lattice Feynmann rules prescribed in this work matches exactly with the conventional approach \cite{hamer}. Note that, our formulation is much more simple as there is no need to deal with any complex 6j coeffiecient \cite{ramesh,hamer} and is well suited for numerical computation.

\section{Summary and Concusions}

In this work, we have used the local loop description in prepotential formulation of lattice gauge theory to construct all possible local gauge invariant operators or linking operators and found their explicit action on all possible local linking states defined locally at each lattice site. We develop a set of `lattice Feynman rules' and hence a complete diagrammatic scheme to perform all computations diagrammatically bypassing long and tedious algebraic calculations. 

The linking number description of local gauge invariant operator and states is over-complete as there exist the Mandelstam constraint. We have solved this constraint explicitly to find all the physical loop configurations consisting of non-intersecting electric flux loops. The physical loop configurations contain nested loops (all non-intersecting) which can overlap with neighbouring loops in one or more segments as shown in figure \ref{physloop}. In order to characterize the physical loop Hilbert space we define a basic loop operator, i.e the smallest plaquette ones which solves the Mandelstam constraint and are a part of the physical loop configuration. We further show that, other configurations can be generated from the basic plaquette loops by applying a set of fusion operators defined on the lattice locally. In fact arbitrary large loops can be generated by local action of these fusion operators. As a consequence of this, the full lattice Hamiltonian is explicitly written in terms of the fusion operators. The complete dynamics of arbitrary non-intersecting loops under this Hamiltonian is thus obtained. 

This diagrammatic tool to handle lattice gauge theories is extremely useful to proceed with lattice calculations analytically in both the strong and weak coupling limit of the theory. The works in these directions, specifically towards the analytic weak coupling expansion is in progress and will be reported shortly. Moreover theses techniques can also find application in numerical simulation of Hamiltonian lattice gauge theories as one can enumerate the complete and physical loop configurations  by just specifying a set of integers locally throughout the lattice without any redundant degrees of freedom and their complete dynamics is already obtained in this work.  

The most novel feature of this approach is that all the steps computed in this work can be performed in any arbitrary dimension, more specifically for $3+1$ dimension which is of physical interest. Addition of Fermions to the theory enlarges the physical configuration space with more local gauge invariant states or linking states but qualitatively the construction steps remain the same. This will be enumerated in a future publication. The recently developed tensor network approach to Hamiltonian lattice gauge theory \cite{tensor1,tensor2} should find this loop formulation most suitable to proceed  with for non Abelian gauge theories. This loop formulation and diagrammatic techniques should also be extremely useful towards the  aim of the construction of quantum simulations \cite{qs} for lattice gauge theories.

\section*{Acknowledgement} 

The authors would like to thank Manu Mathur for many useful informal discussions at multiple stages of this work.
\appendix

\section{Explicit action of loop operators on loop states}

The basic local loop operators arising at a particular site are:
\bea
\hat{\mathcal O}^{i_+j_+} &\equiv &  \frac{1}{\sqrt{( n_i+1)(n_j+2)}}k_+^{ij}\\
\hat{\mathcal O}^{i_+j_-} &\equiv &  \frac{1}{\sqrt{( n_i+1)(n_j+2)}}\kappa^{ij}\\
\hat{\mathcal O}^{i_-j_-} &\equiv &  \frac{1}{\sqrt{( n_i+1)(n_j+2)}}k_-^{ij}
\eea
We now compute the action of these operators on a most general loop state locally characterized by linking numbers as given in (\ref{lij}). Let us first consider the following action: 
\bea
\label{ac++a}
\hat{\mathcal O}^{i_+j_+}|\{l\}\rangle &\equiv &  \frac{1}{\sqrt{( n_i+1)(n_j+2)}}k_+^{ij}|\{l\}\rangle= \frac{(l_{ij}+1)}{\sqrt{( n_i+1)(n_j+2)}}|l_{ij}+1\rangle
\eea
where, $|l_{ij}+1\rangle$ denotes the state in (\ref{lij}) with the particular quantum number $l_{ij}$ increased by 1. This action is simple and straightforward besides being applicable for any $i,j$. We represent the above action pictorially in figure \ref{ac++fig}.

Next we consider,
\bea
\hat{\mathcal O}^{1_+2_-}|\{l\}\rangle &\equiv &  \frac{1}{\sqrt{( n_1+1)(n_2+2)}}\kappa^{21}|\{l\}\rangle \nonumber \\
&=& \frac{1}{\sqrt{( n_1+1)(n_2+2)}}\Bigg[ \frac{\left(k_+^{12}\right)^{l_{12}}\left(k_+^{1\bar 1}\right)^{l_{1\bar 1}}\left(k_+^{1\bar 2}\right)^{l_{1\bar 2}}\left[\kappa^{21},\left(k_+^{2\bar 1}\right)^{l_{2\bar 1}}\right]\left(k_+^{2\bar 2}\right)^{l_{2\bar 2}}\left(k_+^{\bar 1\bar 2}\right)^{l_{\bar 1\bar 2}}}{l_{12}!l_{1\bar 1}!l_{1\bar 2}!l_{2\bar 1}!l_{2\bar 2}!l_{\bar 1\bar 2}!}|0\rangle\nonumber \\
&& + \frac{\left(k_+^{12}\right)^{l_{12}}\left(k_+^{1\bar 1}\right)^{l_{1\bar 1}}\left(k_+^{1\bar 2}\right)^{l_{1\bar 2}}\left(k_+^{2\bar 1}\right)^{l_{2\bar 1}}\left[\kappa^{21},\left(k_+^{2\bar 2}\right)^{l_{2\bar 2}}\right]\left(k_+^{\bar 1\bar 2}\right)^{l_{\bar 1\bar 2}}}{l_{12}!l_{1\bar 1}!l_{1\bar 2}!l_{2\bar 1}!l_{2\bar 2}!l_{\bar 1 \bar 2}!}|0\rangle \Bigg]\\
&=& \frac{1}{\sqrt{( n_1+1)(n_2+2)}}\Bigg[ (l_{1\bar 1}+1)|l_{2\bar 1}-1, l_{1\bar 1}+1\rangle + (l_{1\bar 2}+1)|l_{2\bar 2}-1,l_{1\bar 2}+1\rangle\Bigg]
\eea
In the above calculation we have used the relation:
\bea
k_-(k_+)^p|0\rangle &=& [k_-,(k_+)^p]|0\rangle \nonumber \\
&=& \Bigg[ [k_-,k_+](k_+)^{p-1}+ k_+[k_-,k_+](k_+)^{p-2}+(k_+)^2[k_-,k_+](k_+)^{p-3}+\ldots + (k_+)^{p-1}[k_-,k_+] \Bigg]|0\rangle\nonumber \\
&=& \Bigg[ (\hat n_a+\hat n_b+2)(k_+)^{p-1}+ (\hat n_a+\hat n_b+2-2)(k_+)^{p-1}+(\hat n_a+\hat n_b+2-4)(k_+)^{p-1}+\ldots \nonumber \\ &&+ (\hat n_a+\hat n_b+2-2(p-1))(k_+)^{p-1} \Bigg]|0\rangle\nonumber \\
&=& \Bigg[ (\hat n_a+\hat n_b-2p+4)+(\hat n_a+\hat n_b-2p+6)+\ldots +(\hat n_a+\hat n_b+2) \Bigg](k_+)^{p-1}|0\rangle\nonumber \\
&=& \frac{1}{2}p(2\hat n_a+2\hat n_b+ 6-2p)(k_+)^{p-1}|0\rangle\nonumber \\
&=& p(\hat n_a+\hat n_b+3-p) (k_+)^{p-1}|0\rangle \equiv  p(p+1) (k_+)^{p-1}|0\rangle
\label{comm_tool}
\eea

In general these $\hat{\mathcal O}^{i_+j_-}$ operator acts in the following way:
\bea
\label{ac+-a}
\hat{\mathcal O}^{i_+j_-}|\{l\}\rangle &\equiv &  \frac{1}{\sqrt{( n_i+1)(n_j+2)}}\kappa^{ij}|\{l\}\rangle \nonumber \\
&=& \frac{1}{\sqrt{( n_i+1)(n_j+2)}}\sum_{k\ne i,j} (-1)^{S_{ik}}(l_{ik}+1) |l_{jk}-1,l_{ik}+1\rangle
\eea
where, in any $l_{ij}$ the indices are always ordered in a way such that the first index is always less than the first one, and $$S_{ik}=1~~\mbox{if }i>k~~\&~~S_{ik}=0~~\mbox{if }i<k. $$
We represent the above action pictorially in figure \ref{ac+-fig}.

The last but not the least complicated type of vertex operator is  $\hat{\mathcal O}^{i_-j_-}$ which we calculate using (\ref{comm_tool}). Let's consider the action of the following operator on loop state:
\bea
\hat{\mathcal O}^{1_-2_-}|\{l\}\rangle &\equiv &  \frac{1}{\sqrt{(\hat n_1+2)(\hat n_2+1)}}k_-^{12}|\{l\}\rangle \nonumber \\
&=& \frac{1}{\sqrt{(\hat n_1+2)(\hat n_2+1)}}\Bigg[ \frac{\{ \left[k_-^{12}, \left(k_+^{12}\right)^{l_{12}} \right]+\left(k_+^{12}\right)^{l_{12}}k_-^{12}\} \left(k_+^{1\bar 1}\right)^{l_{1\bar 1}}\left(k_+^{1\bar 2}\right)^{l_{1\bar 2}}\left(k_+^{2\bar 1}\right)^{l_{2\bar 1}}\left(k_+^{2\bar 2}\right)^{l_{2\bar 2}}
\left(k_+^{\bar 1\bar 2}\right)^{l_{\bar 1\bar 2}}}{l_{12}!l_{1\bar 1}!l_{1\bar 2}!l_{2\bar 1}!l_{2\bar 2}!l_{\bar 1\bar 2}!}|0\rangle \Bigg] \nonumber \\
&=& \frac{1}{\sqrt{(\hat n_1+2)(\hat n_2+1)}}\Bigg[ \frac{ l_{12}(n_1+n_2-l_{12}+1)\left(k_+^{12}\right)^{l_{12}-1}\left(k_+^{1\bar 1}\right)^{l_{1\bar 1}}\left(k_+^{1\bar 2}\right)^{l_{1\bar 2}}\left(k_+^{2\bar 1}\right)^{l_{2\bar 1}}\left(k_+^{2\bar 2}\right)^{l_{2\bar 2}}
\left(k_+^{\bar 1\bar 2}\right)^{l_{\bar 1\bar 2}}}{l_{12}!l_{1\bar 1}!l_{1\bar 2}!l_{2\bar 1}!l_{2\bar 2}!l_{\bar 1\bar 2}!}|0\rangle \nonumber \\
&& +\frac{\left(k_+^{12}\right)^{l_{12}}\left(\left[k_-^{12}, \left(k_+^{1\bar 1}\right)^{l_{1\bar 1}}\right] + \left(k_+^{1\bar 1}\right)^{l_{1\bar 1}} k_-^{12} \right)
\left(k_+^{1\bar 2}\right)^{l_{1\bar 2}}\left(k_+^{2\bar 1}\right)^{l_{2\bar 1}}\left(k_+^{2\bar 2}\right)^{l_{2\bar 2}}\left(k_+^{\bar 1\bar 2}\right)^{l_{\bar 1\bar 2}}
}{l_{12}!l_{1\bar 1}!l_{1\bar 2}!l_{2\bar 1}!l_{2\bar 2}!l_{\bar 1\bar 2}!}|0\rangle \Bigg] \nonumber \\
&=& \frac{1}{\sqrt{(\hat n_1+2)(\hat n_2+1)}}\Bigg[(n_1+n_2-l_{12}+1)|l_{12}-1\rangle \nonumber \\ &&+ \frac{\left(k_+^{12}\right)^{l_{12}} l_{1\bar 1} \left(k_+^{1\bar 1}\right)^{l_{1\bar 1}-1}\kappa^{2\bar 1} \left(k_+^{1\bar 2}\right)^{l_{1\bar 2}}\left(k_+^{2\bar 1}\right)^{l_{2\bar 1}}\left(k_+^{2\bar 2}\right)^{l_{2\bar 2}}\left(k_+^{\bar 1\bar 2}\right)^{l_{\bar 1\bar 2}}}{l_{12}!l_{1\bar 1}!l_{1\bar 2}!l_{2\bar 1}!l_{2\bar 2}!l_{\bar 1\bar 2}!}|0\rangle \nonumber \\
&& + \frac{\left(k_+^{12}\right)^{l_{12}}\left(k_+^{1\bar 1}\right)^{l_{1\bar 1}} \left( \left[ k_-^{12}, \left(k_+^{1\bar 2}\right)^{l_{1\bar 2}} \right]+\left(k_+^{1\bar 2}\right)^{l_{1\bar 2}} k_-^{12} \right) \left(k_+^{2\bar 1}\right)^{l_{2\bar 1}}\left(k_+^{2\bar 2}\right)^{l_{2\bar 2}}\left(k_+^{\bar 1\bar 2}\right)^{l_{\bar 1\bar 2}}}{l_{12}!l_{1\bar 1}!l_{1\bar 2}!l_{2\bar 1}!l_{2\bar 2}!l_{\bar 1\bar 2}!}|0\rangle \Bigg]\nonumber \\
&=&  \frac{1}{\sqrt{(\hat n_1+2)(\hat n_2+1)}}\Bigg[ (n_1+n_2-l_{12}+1)|l_{12}-1\rangle \nonumber \\ &&
+ \frac{\left(k_+^{12}\right)^{l_{12}} l_{1\bar 1} \left(k_+^{1\bar 1}\right)^{l_{1\bar 1}-1} \left(k_+^{1\bar 2}\right)^{l_{1\bar 2}}\left(k_+^{2\bar 1}\right)^{l_{2\bar 1}}\left[\kappa^{2\bar 1},\left(k_+^{2\bar 2}\right)^{l_{2\bar 2}}\right]\left(k_+^{\bar 1\bar 2}\right)^{l_{\bar 1\bar 2}}}{l_{12}!l_{1\bar 1}!l_{1\bar 2}!l_{2\bar 1}!l_{2\bar 2}!l_{\bar 1\bar 2}!}|0\rangle \nonumber \\ 
&& +  \frac{\left(k_+^{12}\right)^{l_{12}}\left(k_+^{1\bar 1}\right)^{l_{1\bar 1}} l_{1\bar 2} \left(k_+^{1\bar 2}\right)^{l_{1\bar 2}-1}\kappa^{2\bar 2} \left(k_+^{2\bar 1}\right)^{l_{2\bar 1}}\left(k_+^{2\bar 2}\right)^{l_{2\bar 2}}\left(k_+^{\bar 1\bar 2}\right)^{l_{\bar 1\bar 2}}}{l_{12}!l_{1\bar 1}!l_{1\bar 2}!l_{2\bar 1}!l_{2\bar 2}!l_{\bar 1\bar 2}!}|0\rangle \Bigg]\nonumber \\
&=& \frac{1}{\sqrt{(\hat n_1+2)(\hat n_2+1)}}\Bigg[ (n_1+n_2-l_{12}+1)|l_{12}-1\rangle + (l_{\bar 1\bar 2}+1)(-1)^{S_{\bar 1\bar 2}} |l_{1\bar 1}-1,l_{2\bar 2}-1,l_{\bar 1\bar 2}+1\rangle \nonumber \\ &&
+ (l_{\bar 1\bar 2}+1)(-1)^{S_{\bar 2\bar 1}} |l_{1\bar 2}-1,l_{2\bar 1}-1,l_{\bar 1\bar 2}+1\rangle \Bigg]
\eea
with, $$S_{ik}=1~~\mbox{if }i>k~~\&~~S_{ik}=0~~\mbox{if }i<k. $$

Hence, for a general $\hat{\mathcal O}^{i_-j_-}$ operator, the action is:
\bea
\label{ac--a}
\hat{\mathcal O}^{i_-j_-}|\{l\}\rangle = \frac{1}{\sqrt{(\hat n_i+2)(\hat n_j+1)}}\Bigg[ (n_i+n_j-l_{ij}+1)|l_{ij}-1\rangle + \sum_{\bar i,\bar j\{\ne i,j\}}(l_{\bar i \bar j}+1)(-1)^{S_{\bar i\bar j}}| l_{i\bar i}-1,l_{j\bar j}-1,l_{\bar i \bar j}+1\rangle  \Bigg]
\eea
We represent the above action pictorially in figure \ref{ac--fig}.

\newpage
\section{Normalization of the Loop States}

The linking states at a particular site of a two dimensional spatial lattice, are characterized by six linking numbers $l_{12},l_{1\bar 1},l_{1\bar 2},l_{2\bar 1},l_{2 \bar 2},l_{\bar 1\bar 2}$. The SU(2) flux along each directions at a particular site are counted as in (\ref{n1},\ref{n2},\ref{n3},\ref{n4}).
Moreover there exists the  Mandelstam constraint given in (\ref{mcsol}), which must be solved in order to get independent loop states  implying that at each site $x$, 
 atleast either of the two quantum numbers $l_{1\bar 1},l_{2\bar 2}$ must be zero.
Hence, after solving the Mandelstam constraint, only five non-zero linking quantum number together with the two Abelian constraints are present at each site. \\ 
Any linking state, characterized by five non-zero linking number is always orthogonal with respect to the four number operators defined in (\ref{n1})-(\ref{n4}) but there exists a fifth quantum number which makes the orthogonality non-trivial as given below:
\bea
\langle \{l'_{ij}\} |\{l_{ij}\} \rangle = \prod_{i=1,2,\bar 1\bar 2}\delta_{n'_i,n_i} F(\{l'_{ij}\},\{l_{ij}\})
\eea
Before determining the complicated function $F(\{l'_{ij}\},\{l_{ij}\})$, let us first realize the orthogonality of linking states in terms of four quantum numbers. This can be realized trivially when one consider the linking state which has only four non-zero linking number, such as for example with $l_{12}=0$, besides $l_{1\bar 1}(x)l_{2\bar 2}(x)=0$. The orthonormality of such states are obtained as:
\bea
\langle l'_{12}=0|l_{12}=0\rangle &=& \langle l'_{1\bar 2},l'_{2\bar 1},l'_{\bar 1\bar 2},l'_{1\bar 1},l'_{2\bar 2}| l_{1\bar 2},l_{2\bar 1},l_{\bar 1\bar 2},l_{1\bar 1},l_{2\bar 2}\rangle \nonumber \\
&=& \frac{\left(l_{1\bar 2}+l_{2\bar 1}+ l_{\bar 1\bar 2}+l_{1\bar 1}+l_{2\bar 2}+1 \right)!}{l_{1\bar 2}!\left( l_{2\bar 1}+l_{\bar 1\bar 2}+l_{1\bar 1}+l_{2\bar 2}+1 \right)!} \delta_{l'_{1\bar 2},l_{1\bar 2}} \nonumber \\
&& \times\frac{\left(l_{2\bar 1}+ l_{\bar 1\bar 2}+l_{1\bar 1}+l_{2\bar 2}+1 \right)!}{l_{2\bar 1}!\left( l_{\bar 1\bar 2}+l_{1\bar 1}+l_{2\bar 2}+1 \right)!} \delta_{l'_{2\bar 1},l_{2\bar 1}} \nonumber \\
&& \times\frac{\left( l_{\bar 1\bar 2}+l_{1\bar 1}+l_{2\bar 2}+1 \right)!}{l_{\bar 1\bar 2}!\left( l_{1\bar 1}+l_{2\bar 2}+1 \right)!} \delta_{l'_{\bar 1\bar 2},l_{\bar 1\bar 2}} \nonumber \\
&& \times\left(l_{1\bar 1}+1 \right)\left(l_{2\bar 2}+1 \right) \delta_{l'_{1\bar 1},l_{1\bar 1}}  \delta_{l'_{2\bar 2},l_{2\bar 2}} \nonumber \\
&\equiv & B_p \, \delta_{l'_{1\bar 2},l_{1\bar 2}} \delta_{l'_{2\bar 1},l_{2\bar 1}}\delta_{l'_{\bar 1\bar 2},l_{\bar 1\bar 2}}\delta_{l'_{1\bar 1},l_{1\bar 1}} \delta_{l'_{2\bar 2},l_{2\bar 2}}\label{b}
\eea
(\ref{b}) is obtained by extracting the $k_+^{ij}$ operator from the bra state and acting that on the ket state following (\ref{ac--}) until it reaches $l'_{ij}=0$ for all nonzero $l_{ij}$, considering one by one.

The next complicated orthogonality arises when either of the bra and ket state has $5$ non-zero linking numbers and the other one has only $4$. For example, consider the following case:
\bea
\langle l'_{12}=0| \{l_{ij}\}\rangle &=& \frac{1}{l_{12}}\langle l'_{12}=0| k_+^{12}|l_{12}-1\rangle  \nonumber \\
&=& \frac{1}{l_{12}(l_{12}-1)}\left[0-(l'_{\bar 1\bar 2}+1) \langle l'_{12}=0,l'_{1\bar 2}-1,l'_{2\bar 1}-1,l'_{\bar 1\bar 2}+1| k_+^{12}|l_{12}-2\rangle\right]\nonumber \\
&=& A_1^{'(1)}\langle l'_{12}=0,l'_{1\bar 2}-1,l'_{2\bar 1}-1,l'_{\bar 1\bar 2}+1| k_+^{12}|l_{12}-2\rangle\nonumber \\
&=& \vdots\nonumber \\
&& \vdots\nonumber \\
&=& A_1^{'(1)} A_1^{'(2)}\ldots A_1^{'(l_{12})} \langle l'_{12}=0,l'_{1\bar 2}-l_{12},l'_{2\bar 1}-l_{12},l'_{\bar 1\bar 2}+l_{12}|l_{12}=0\rangle
\label{a'}
\eea
where, \bea A_1^{'(i)}=-\frac{l'_{\bar 1\bar 2}+i}{l_{12}+i-1}.\label{a'exp}\eea
(\ref{a'}) is also obtained by extracting the $k_+^{12}$ operator from the ket state and acting that on the bra state following (\ref{ac--}) until it reaches $l_{12}=0$. The orthogonality of the final state in (\ref{a'}) is already given in (\ref{b}).

Now moving further towards the most complicated and general situation where both the bra and ket states has five non-zero linking numbers, the orthogonality of that state is again obtained in terms of the already calculated orthonormal states in (\ref{a'}) and (\ref{b}). Let us consider the orthogonal linking loop state defined at a site $x$, characterized by the set of $5$ linking numbers as follows 
$$ |l_{12},l_{1\bar 2},l_{2\bar 1},l_{\bar 1\bar 2},( l_{1\bar 1}/l_{2\bar 2})\rangle $$
These loop states are trivially orthogonal with respect to $n_i$'s for $i=1,2,\bar 1,\bar 2$, but non-trivial  orthonormality exists in terms of the linking quantum number.
 To calculate the orthogonality of loop states in terms of the linking numbers 
 , we take an iterative approach as discussed below:
 Let us consider the following arbitrary overlap of the states
\bea
\langle \{ l'_{ij} \}|\{ l_{ij} \}\rangle = \frac{1}{l'_{12}}\langle l'_{12}-1| k_-^{12}|\{ l_{ij} \}\rangle
\label{step1}
\eea
Note that, in the right hand side of the above equation, in both the bra and ket states we have mentioned the linking number, only which has been changed. We will maintain this approach in the later part of this section as well by characterizing a newly produced state by the changed linking numbers only. Whenever, none of the linking numbers do change, we will characterize the state by the whole set of linking numbers $\{l_{ij}\}$.
Now from the action given in  (\ref{ac--}) on the loop states which satisfies Mandelstam constraint (\ref{mcsol}), one obtain 
\bea
 k_-^{12}|\{ l_{ij} \}\rangle &=&
  (n_1+n_2-l_{12}+1)|l_{12}-1\rangle - (l_{\bar 1\bar 2+1})|l_{1\bar 2}-1,l_{2\bar 1}-1,l_{\bar 1\bar 2}+1\rangle 
\eea
with $n_1=l_{12}+l_{1\bar 1}+l_{1\bar 2}$ and $n_2=l_{12}+l_{2\bar 1}+l_{2\bar 2}$. Note that, in the right hand side of the above equation we have suppressed the quantum numbers which remain unchanged. 
In this way, as done in (\ref{step1}), one can extract out a particular $k_-{ij}$ operator from the bra state or $k_+{ij}$ from the ket state state and act that on the corresponding ket/bra state to increase or decrease the $l_{ij}$ quantum numbers by one unit until that particular $l_{ij}$ or $l'_{ij}$ is exhausted. Or in other way, the iteration can stop at a certain value of $l_{ij}$, (for example $l_{1\bar 2}$ and $l_{2\bar 1}$ as shown in the above example) which is being decreased by one unit for each step of the iterations. Hence, clearly iteration will continue  $p$ times, where $p= \min{(l_{12},l_{1\bar 2},l_{2\bar 1},l_{\bar 1\bar 2},l'_{12},l'_{1\bar 2},l'_{2\bar 1},l'_{\bar 1\bar 2})}$. Continuing with the example discussed above in (\ref{step1}) and considering the Mandelstam constraint at that particular site by putting $l_{1\bar 1}=0$, we finally get:
\bea
&& \langle\{ l'_{ij} \}|\{ l_{ij} \}\rangle \nonumber \\ &=& \frac{(l_{12}+l_{1\bar 2}+l_{2\bar 1}+l_{1\bar 1}+l_{2\bar 2}+1)}{l'_{12}} \langle l'_{12}-1 |l_{12}-1\rangle\nonumber \\ &&  - \frac{(l_{\bar 1\bar 2}+1)}{l'_{12}} \langle l'_{12}-1  |l_{1\bar 2}-1,l_{2\bar 1}-1,l_{\bar 1\bar 2}+1\rangle\nonumber \\
&\equiv & A_0^{(1)}  \langle l'_{12}-1 |l_{12}-1\rangle 
 + A_1^{(2)}  \langle l'_{12}-1 |l_{1\bar 2}-1,l_{2\bar 1}-1,l_{\bar 1\bar 2}+1\rangle \nonumber \\
&&\mbox{(Repeating one more step of iteration for the two overlaps separately,)}\nonumber \\
&& \nonumber\\
&=& A_0^{(1)}\Bigg[   A_0^{(2)}  \langle l'_{12}-2 |l_{12}-2\rangle 
+ A_1^{(2)}  \langle l'_{12}-2 |l_{12}-1,l_{1\bar 2}-1,l_{2\bar 1}-1,l_{\bar 1\bar 2}+1\rangle   \Bigg]\nonumber \\
\nonumber \\ &+& 
A_1^{(1)} \Bigg[   A_0^{(2)}  \langle l'_{12}-2 |l_{12}-1,l_{1\bar 2}-1,l_{2\bar 1}-1,l_{\bar 1\bar 2}+1\rangle + A_1^{(2)} \langle l'_{12}-2 |l_{12}-1,l_{1\bar 2}-2,l_{2\bar 1}-2,l_{\bar 1\bar 2}+2\rangle \Bigg]\nonumber \\ 
&\equiv& A_0^{(1)}   A_0^{(2)}\langle l'_{12}-2 |l_{12}-2\rangle \nonumber \\ && 
+\left[ A_0^{(1)}   A_1^{(2)} +A_1^{(1)}    A_0^{(2)} \right]\langle l'_{12}-2 |l_{12}-1,l_{1\bar 2}-1,l_{2\bar 1}-1,l_{\bar 1\bar 2}+1\rangle   \nonumber \\&& + A_1^{(0)} A_1^{(1)}   
\langle l'_{12}-2 |l_{12}-1,l_{1\bar 2}-2,l_{2\bar 1}-2,l_{\bar 1\bar 2}+2\rangle
 \nonumber \\
&=& \vdots \nonumber \\ && \vdots\nonumber \\
&& \mbox{(After $p^{th}$  iteration, for example if $p=l'_{12}$)}\nonumber \\
&& \nonumber\\
&\equiv & \sum_{q=0}^p\left[ \sum^{~~~~~\prime}_{\{s_i \}_q} \left(A_{s_1}^{(1)}A_{s_2}^{(2)}\ldots A_{s_p}^{(p)}\right)\langle l'_{12}=0|l_{12}-p+q,l_{1\bar 2}-q,l_{2\bar 1}-q,l_{\bar 1\bar 2}+q\rangle\right]\label{geno}
\eea
where, each $s_i$ can take values of either $1$ or $0$, and  the $\sum^{~~~~~\prime}_{\{s_i \}_q}$ denotes that the sum is over all permutations of the set $$ \{s_i\}_q\equiv P\left( \underbrace{1,1,\ldots,1}_{q~\mbox{times}},\underbrace{0,0,\ldots,0}_{p-q~\mbox{times}}\right) $$ The coefficients $A^{(i)}_{s_i}$'s are given by,
\bea
&& A_0^{(1)} = \frac{(l_{12}+l_{1\bar 2}+l_{2\bar 1}+l_{1\bar 1}+l_{2\bar 2}+1)}{l'_{12}}~~ ~~~~~,~~~~~  A_1^{(1)} = -\frac{(l_{\bar 1\bar 2}+1)}{l'_{12}} \nonumber \\
&& A_0^{(2)} = \frac{(l_{12}+l_{1\bar 2}+l_{2\bar 1}+l_{1\bar 1}+l_{2\bar 2})}{(l'_{12}-1)} ~~~~~,~~~~~  A_1^{(2)} = -\frac{(l_{\bar 1\bar 2}+2)}{(l'_{12}-1)}\nonumber \\
& &~~~~~~~~~~~~~~~~~~~~~~~~~~~~~~~~~~~~~~~~~~~~~~~~~~~~~\vdots \nonumber \\
&& A_0^{(p)} = \frac{(l_{12}+l_{1\bar 2}+l_{2\bar 1}+l_{1\bar 1}+l_{2\bar 2}+2-p)}{(l'_{12}-p+1)} ~~~~~,~~~~~  A_1^{(p)} = -\frac{(l_{\bar 1\bar 2}+p)}{(l'_{12}-p+1)}\nonumber \\
\eea
In this particular example, the iteration stops at $p^{th}$ level as the at the final step contain the overlap given below,
\bea
\langle l'_{12}=0| l_{12}-p+q,l_{1\bar 2}-q,l_{2\bar 1}-q,l_{\bar 1\bar 2}+q\rangle
\eea
Clearly, the ket state contain four nonzero $l'_{ij}$'s whereas the bra state has five, the norm of which is given in (\ref{a'}) in terms of the norm given in (\ref{b}).
Using these, for our example case, after a few steps of algebra we have,
\bea
\langle l'_{12}=0| l_{12}-p+q,l_{1\bar 2}-q,l_{2\bar 1}-q,l_{\bar 1\bar 2}+q\rangle 
&=& \tilde A_q^{'(1)} \tilde A_q^{'(2)}\ldots \tilde A_q^{'(l_{12}-l'_{12})} \tilde B^q_{l_{12}}~\nonumber \\ &&\delta_{l'_{1\bar 2}+l'_{12},l_{1\bar 2}+l_{12}} \delta_{l'_{2\bar 1}+l'_{12},l_{2\bar 1}+l_{12}}\delta_{l'_{\bar 1\bar 2}-l'_{12},l_{1\bar 2}-l_{12}}\delta_{l'_{1\bar 1},l_{1\bar 1}}\delta_{l'_{2\bar 2},l_{2\bar 2}}~~~~~~~~~~~~~~\label{fo}
\eea
where , \bea\tilde A_q^{'i}= -\frac{l'_{\bar 1\bar 2}+i}{l_{12}-l'_{12}+q+i-1}, ~~\mbox{for }i=1,2,\ldots, l_{12}-l'_{12}+q \label{ao}\eea is obtained using (\ref{a'exp}) and
\bea
\tilde B^q_{l_{12}}&=& \frac{\left(l_{1\bar 2}+l_{2\bar 1}+ l_{\bar 1\bar 2}+l_{1\bar 1}+l_{2\bar 2}+1 -q\right)!}{(l_{1\bar 2}-q)!\left( l_{2\bar 1}+l_{\bar 1\bar 2}+l_{1\bar 1}+l_{2\bar 2}+1 \right)!}  \times\frac{\left(l_{2\bar 1}+ l_{\bar 1\bar 2}+l_{1\bar 1}+l_{2\bar 2}+1 \right)!}{(l_{2\bar 1}-q)!\left( l_{\bar 1\bar 2}+l_{1\bar 1}+l_{2\bar 2}+1+q \right)!} \nonumber \\
&& \times\frac{\left( l_{\bar 1\bar 2}+l_{1\bar 1}+l_{2\bar 2}+1+q \right)!}{(l_{\bar 1\bar 2}+q)!\left( l_{1\bar 1}+l_{2\bar 2}+1 \right)!}  \times\left(l_{1\bar 1}+1 \right)\left(l_{2\bar 2}+1 \right)\label{bo}
\eea
is obtained using (\ref{b}) for our case.

 Hence, the complete orthonormality relation of the states $|\{l_{ij}\}\rangle$ as our example with $p=l'_{12}$, can be obtained combining the (\ref{geno}) and (\ref{fo}) as,
\bea
&& \sum_{q=0}^p\left[\left[ \sum^{~~~~~\prime}_{\{s_i \}_q} \left(A_{s_1}^{(1)}A_{s_2}^{(2)}\ldots A_{s_p}^{(p)}\right)\right]\frac{(-1)^{l_{12}-p+q}(l'_{\bar 1\bar 2}+l_{12}-p+q)!}{l'_{\bar 1\bar 2}!(l_{12}-p+q)!}\tilde B_{p}^q\right] 
\delta_{l'_{1\bar 2}+p,l_{1\bar 2}+l_{12}} \delta_{l'_{2\bar 1}+p,l_{2\bar 
 1}+l_{12}} \nonumber \\ && ~~~~ \delta_{l'_{\bar 1\bar 2}-p,l_{\bar 1\bar 2}-l_{12}}\delta_{l'_{1\bar 1},l_{1\bar 1}}\delta_{l'_{2\bar 2},l_{2\bar 2}} 
\eea
where,  $\tilde B^q_{p}$ are defined in (\ref{bo}).

Moving away from this particular example, the most general case can have any of the $l_{ij}$'s as minimum and the same calculation will go through. The final expression of any arbitrary case (i.e for any arbitrary p) can be easily read off from the expression derived above just by replacing the role of $l_{12}/l'_{12}$ by the corresponding $p$.

\section{Strong Coupling Perturbation Expansion}

The unperturbed Hamiltonian in the limit $g\rightarrow 0$ is the electric part of the Hamiltonian $H_{e}$ given in (\ref{ham1}). $H_{e}$ is solved exactly yielding the loop states as the strong coupling eigenstates with eigenvalues measuring the total flux around the loop. The strong coupling vacuum satisgying $H_e|0\rangle =0$ is the state with no loop present and has unperturbed energy eigenvalue or the unperturbed vacuum energy $E_0^{(0)}=0$. We now calculate perturbative corrections to this vacuum energy for the first couple of orders analytically.  Rayleigh-Schrödinger perturbation theory gives the corrections to the vacuum energy as:
\bea
E_0= E_0^{(0)}+\frac{1}{g^2} E_0^{(1)}+\frac{1}{g^4} E_0^{(2)}+\frac{1}{g^6} E_0^{(3)}+\frac{1}{g^8} E_0^{(4)}+\ldots\ldots
\eea
The first order correction is given by $\langle 0|H_{I}|0\rangle=0$ for $H_I=H_{mag}$. Similarly all odd orders of corrections to vacuum energy do vanish implying the full correction to come only from even orders. 
The lowest order correction is of second order and is given by,
\bea
E_0^{(2)} &=& \sum_{n_1\ne 0} \frac{\langle 0|H_I|n_1\rangle \langle n_{1}|H_I|0\rangle}{\langle n_1|n_1\rangle\left(E_0-E_0^{n_1}\right)} = \sum_{n_1\ne 0} \frac{|\langle n_{1}|H_I|0\rangle|^2}{\langle n_1|n_1\rangle\left(E_0-E_0^{n_1}\right)}  \label{e02} 
\eea
where, $H_I\equiv H_{mag}=2\mbox{Tr} U_{plaquette}$ for SU(2) case. In (\ref{e02}), $|n_1\rangle$ is always the state created by a single action of $\mbox{Tr} U_{plaquette}$ on $|0\rangle$, and it can only be a single plaquette state created by the first term $H_1$ of the 16 terms figure \ref{h16} on vacuum. Obviously for a latice consisting of $N$ number of plaquettes, there exists $N$ such $|n_1\rangle$ states which contributes to the perturbation expansion of vacuum energy.  Note that, each of the loops contributing to the perturbation expansion which are eigenstates of the unperturbed Hamiltonian has its unperturbed energy given by,
\bea
H_{el}|n_i\rangle = \sum_{links} E^2_{links} |n_i\rangle = \sum_{links} \frac{n}{2}\left(\frac{n}{2}+1\right) |n_i\rangle \forall i
\eea
for a loop state with, $n$  units of flux along a particular link. For example the single plaquette states $|n_1\rangle$ will have $E_0^{n_1}=4\times \frac{3}{4}=3$. Hence, the second order correction is finally obtained as,
\bea
E_0^{(2)}= N \frac{|\langle L(\tilde x)=1|2\mbox{Tr} U_{plaquette}|0\rangle|^2}{\langle L(\tilde x)=1|L(\tilde x)=1\rangle}\times \frac{1}{\left(-4\times \frac{3}{4}\right)}=N\times  \frac{2^2}{-3}
\eea
Note that, this final result is obtained after using the action of $H_{mag}$ as obtained in the earlier sections and the normalization of the state is obtained using Appendix B. This correction matches exactly\footnote{upto a factor of $2^2$, which is due to the mismatch of the Hamiltonian in (\ref{ham1}) and that in \cite{hamer}. } to the correction in \cite{hamer} for this order. 
To confirm the viability of our formulation, we further proceed to calculate the next order correction given by
\bea
E_0^{(4)} &=& \sum_{\{n_i\}_{\ne 0}} \frac{\langle 0|H_I|n_1\rangle\langle n_1|H_I|n_2\rangle  \langle n_2|H_I|n_3\rangle   \langle n_3|H_I|0\rangle}{\langle n_1|n_1\rangle\langle n_2|n_2\rangle\langle n_3|n_3\rangle\left(E_0-E_0^{n_1}\right)\left(E_0-E_0^{n_2}\right)\left(E_0-E_0^{n_3}\right)} \nonumber\\&&  - E_0^{(2)} \sum_{\{n_1\}_{\ne 0}} \frac{\langle 0|H_I|n_1\rangle\langle n_1|H_I|0\rangle}{\langle n_1|n_1\rangle\left(E_0-E_0^{n_1}\right)^2}\label{e04}
\eea
Note that, in the fourth order corrections $|n_1\rangle$ as well as the $|n_3\rangle$ are the single plaquette states, located anywhere on the lattice.
 $E_0^{(4)}$ involves another intermediate state $|n_2\rangle$ which is a two plaquette state. Now there exists the following possibility for the two plaquette states:
\begin{enumerate}
\item $|n_2\rangle = H_1|n_1\rangle\equiv |L(\tilde x_1)=1, L(\tilde x_2)=1\rangle$, i.e two decoupled plaquette loops located anywhere in the lattice without any overlap or touch with the first plaquette. Clearly for each $|n_1\rangle$, there are $N-9$ possible $|n_2\rangle$ with $E_0^{n_2}=8\times\frac{1}{2}\left(\frac{1}{2}+1\right)=6$.
\item The second plaquette can be created by the action of $H_1$ but with complete overlap with the first one, i.e $|n_2\rangle \equiv |L(\tilde x)=2\rangle$. In this case, $E_0^{n_2}=4\times\frac{2}{2}\left(\frac{2}{2}+1\right)=8$. The norm of such state can be calculated from Appendix B.
\item There exists four possibilities of the two plaquette state to be two separate plaquettes with overlap along any of the link, i.e $|n_2\rangle = H_1|n_1\rangle\equiv |L(\tilde x_1)=1, L(\tilde x_1\pm e_1(\pm e_2))=1\rangle$ with $E_0^{n_2}=\frac{2}{2}\left(\frac{2}{2}+1\right)+ 6\times \frac{1}{2}\left(\frac{1}{2}+1\right)=\frac{13}{2}$ and respective norms.
\item The second plaquette can again be created by $H_1$ in four other possible ways, where the two plaquettes are touching each other at one of its four corners, i.e $|n_2\rangle = H_1|n_1\rangle\equiv |L(\tilde x_1)=1, L(\tilde x_1\pm e_1\pm e_2)=1\rangle$. For those states $E_0^{n_2}=6$, but norm is different and can be calculated easily.
\item By the action of type (b) terms in the Hamiltonian, the two plaquette state can be a loop carrying unit flux with verical extension of two lattice units and horizontal extension of one, i.e  $|n_2\rangle = H_{3/5}|n_1\rangle\equiv |L(\tilde x_1)=1, L(\tilde x_1\pm e_2)=1, N_1(\tilde x\pm \frac{e_2}{2})=1\rangle$. These two states are with $E_0^{n_2}=6\times\frac{1}{2}\left(\frac{1}{2}+1\right)=\frac{9}{2}$ and with certain norm.
\item Similarly, by the action of type (b) terms in the Hamiltonian, the two plaquette state can be a loop carrying unit flux with verical extension of one lattice units and horizontal extension of two, i.e  $|n_2\rangle = H_{2/4}|n_1\rangle\equiv |L(\tilde x_1)=1, L(\tilde x_1\pm e_1)=1, N_2(\tilde x\pm \frac{e_1}{2})=1\rangle$  with $E_0^{n_2}=\frac{9}{2}$ and norm to be caculated from Appendix B.
\end{enumerate}
Explicit calculation incorporating all the coefficients given in table 1 for the Hamiltonian actions and the norm of each state calculated using the appendix we finally obtain,
\bea
E_0^{(4)} = N\frac{2\times 163}{3^4\times 13}\equiv N\times 2^4\times \frac{163}{8424}
\eea
At this order also the result matches exactly (i.e upto 12th decimal place)\footnote{upto a factor of $2^4$, which is due to the mismatch of the Hamiltonian in (\ref{ham1}) and that in \cite{hamer}. } with \cite{hamer}.
In the same way the strong coupling perturbation correction to any loop state can be performed within this scheme and note that this scheme is independent of any cluster size or lattice size.

Besides making strong coupling perturbation expansion viable upto any arbitrary order our formulation is also suitable to approach towards weak coupling limit. It seems that the fusion variables become extremely important to work with in this regime. The work in this direction is in progress and will be reported shortly.

\end{document}